\documentclass[aps,prb,twocolumn,amsmath,amssymb,reprint,groupedaddress]{revtex4-1}

\usepackage{float}
\usepackage{graphicx}  
\usepackage{xcolor}
\usepackage{dcolumn}   
\usepackage{bm}        
\usepackage{amssymb}   
\usepackage{epstopdf}
\usepackage{xfrac}
\usepackage{subfigure}
\usepackage{anyfontsize}

\usepackage{amsmath}
\usepackage{amsfonts}
\usepackage{bbm}
\definecolor{darkblue}{rgb}{0.1,0.2,0.6}
\definecolor{darkred}{rgb}{0.8,0.1,0.2}
\usepackage[colorlinks,citecolor=darkblue,linkcolor=darkred,urlcolor=darkblue]{hyperref} 
\usepackage{tikz}
\usepackage{enumerate}
\usepackage{setspace}
\usepackage{url}  

\newcommand{\del}{\nabla}
\numberwithin{equation}{section}
\renewcommand\theequation{\arabic{section}.\arabic{equation}}

\begin{document}

\title{Viscoelastic response of topological tight-binding models in two and three dimensions}
\author{Hassan Shapourian}
\author{Taylor L. Hughes}
\author{Shinsei Ryu}

\affiliation{Department of Physics, University of Illinois at Urbana-Champaign, Urbana Illinois 61801, USA}

\begin{abstract}
The topological response to external perturbations is an effective probe to characterize different topological phases of matter. Besides the Hall conductance, the Hall viscosity is another example of such a response that measures how electronic wave functions respond to changes in the underlying geometry. 
Topological (Chern) insulators are known to have a quantized Hall conductance. A natural question is how the Hall viscosity behaves for these materials.
So far, most of studies on the Hall viscosity of Chern insulators have focused on the continuum limit.
The presence of lattice breaks the continuous translational symmetry to a discrete group and this causes two complications: it introduces a new length scale associated with the lattice constant, and makes the momentum periodic. We develop two different methods of how to implement a lattice deformation: (1) a lattice distortion is encoded as a shift in the lattice momentum, and (2) a lattice deformation is treated microscopically in the gradient expansion of the hopping matrix elements. After establishing the method of deformation we can compute the Hall viscosity through a linear response (Kubo) formula. We examine these methods for three models: the Hofstadter model, the Chern insulator, and the surface of a 3D topological insulator. Our results in certain regimes of parameters, where the continuum limit is relevant, are in agreement with previous calculations. We also provide possible experimental signatures of the Hall viscosity by studying the phononic properties of a single crystal 3D topological insulator.
\end{abstract}

\maketitle

\section{Introduction}


Topological phases are unusual phases of matter which cannot be categorized using Landau's symmetry-breaking paradigm\cite{wenbook,fradkinbook}. The recent discovery of topological insulators (TI) has opened new avenues for the study of symmetry-protected topological phases\cite{ZhangRev,KaneRev}, which require some additional symmetries in order to remain in a robust topological phase. In all of these cases, the topological phase is protected by a bulk energy gap, e.g., in the case of the quantum Hall effects (QHE), the gap and non-trivial topology are due to an interplay of strong magnetic field and possibly electron-electron interactions. While topological phases have provided a fertile ground for theoretical study, there are also many experimental examples of various topological phases\cite{Klitzing1980,*Tsui1982, Molenkamp2007,*Hasan2008,*Chen10072009,*XiaBS_exp}.

Since topological phases do not have a local order parameter,  one of the key goals for describing topological phases is to find a minimal set of quantities that fully characterizes each distinct phase. Such quantities provide a framework to classify topological phases, and more importantly, may be related to the physical response coefficients that can be measured experimentally, or used to distinguish topological phases in numerical simulations. A canonical example is the quantized Hall conductance for the QHE\cite{TKNN1982,Laughlin_spec,PhysRevB.31.3372, Klitzing1980,*Tsui1982}, while another example is the topological magneto-electric effect for 3D time-reversal invariant TIs\cite{Qi_3DTI,Moore-Vanderbilt}.

Here, we are interested in another example of a response coefficient called the Hall viscosity~\cite{Avron1995}. This viscosity response is one part of the electronic viscoelastic stress response to an applied strain $u_{ij}$:
\begin{align*}
\langle \hat{\cal T}_{ij}\rangle = \Lambda_{ijkl} u_{kl} + \zeta_{ijkl} \dot{u}_{kl},
\end{align*}
where $\hat{\cal T}$ is the stress tensor, $\Lambda$ and $\zeta$ are the elasticity and viscosity tensors respectively, and the strain $u_{ij}=\tfrac{1}{2}(\partial_i u_j+\partial_j u_i)$ is a symmetrized gradient of the displacement $u_{i}.$ 
The Hall viscosity is one contribution to $\zeta$ and is anti-symmetric in exchanging the first $ij$ and second $kl$ pairs of indices, and hence non-dissipative. For our purposes, we only consider isotropic systems (or systems with at least $C_4$ rotation symmetry), where the only non-zero components of the Hall-viscosity response are $\zeta_{1112}=\zeta_{1222}$, which we denote by $\zeta_H$.
Similar to the Hall conductance, $\zeta_H$ can be derived from an adiabatic response calculation and is completely quantum in nature; it also requires time-reversal symmetry to be broken to be non-vanishing~\cite{Avron1995}. 

The Hall viscosity was originally studied in the integer and fractional QHE~\cite{Avron1995,Levay1995,Avron_multi,Read_nonAb,Haldane,Read_Rezayi,Read_Kubo,Hoyos,DTSon_eff,Tokatly,DTSon_AdsCFT,PhysRevB.85.184503,Maissam2012,Wiegmann12,
PhysRevB.85.045104,DMRG_Hvisc,Hidaka2013,Rudro,Abanov13,PhysRevB.90.115139,PhysRevB.89.125303,Hoyos_rev,PhysRevB.91.125303,
PhysRevB.90.014435,PhysRevLett.113.266802,PhysRevLett.113.046803,PhysRevLett.114.016802,PhysRevB.90.045123,
PhysRevLett.114.016805,PhysRevB.91.035122} and, for the integer effect, was shown to be $\zeta_H= \hbar \nu^2 /8\pi \ell_B^2$, where $\nu= nh/eB$ is the filling fraction, and $\ell_B= (\hbar/|eB|)^{1/2}$ is the magnetic length. 
The viscoelastic response of the Chern insulator, a model for the integer QHE without Landau levels\cite{Haldane_CI}, has also been investigated using a massive Dirac model~\cite{Taylor2011}, and it has been shown that the Hall viscosity in the non-trivial Chern insulator phase is of the form $\zeta=\hbar/8\pi \ell^2$, where the emergent length scale depends on the bulk mass gap, $m$, and the Fermi velocity $v_F$, as $\ell=\hbar v_F/2m$.
In both the integer QHE and Chern insulator one can write a continuum field theory from which the Hall viscosity is calculated as a response to deformations in the underlying geometry of the system. In the former, the magnetic length $\ell_B$ introduces a minimal length/area scale, and $\zeta_H$ is inherently finite; while in the latter, a regularization scheme was proposed to obtain the finite result above\cite{Taylor2011,Taylor2013}. There have also been some recent studies of Hall viscosity response in 3D as well\cite{Onkar,visc_Weyl}. 

One open problem is the calculation of the Hall viscosity in lattice models away from the continuum limit. For example,  in a discrete lattice tight-binding model it is not clear, \emph{a priori}, how to precisely model the geometric deformations needed to generate a Hall viscosity response.
In this article, we attack this problem by developing and comparing two different numerical methods to compute the Hall viscosity in tight-binding lattice models. These methods differ in the way geometric deformations are treated in the lattice model, and represent two natural lattice geometry modifications. Our first method is inspired by the analogy between the viscoelastic and electromagnetic responses, where we essentially introduce a frame field Peierls factor. 
In the second method, we use a more physical approach for how the strain can be realized by imagining mechanically moving the lattice degrees of freedom from their equilibrium positions and calculating the modifications to tight-binding overlap integrals.

There have been some earlier pioneering works on Hall viscosity calculations in lattice models. First, the idea of coupling the electronic structure to phononic degrees of freedom, along the same lines as our second method, was originally studied by Barkeshli {\it et al.}~\cite{Maissam2012}, where they proposed the ``phonon Hall viscosity" as the adiabatic response of the electron state to acoustic phonons. They estimated the corrections to the phononic dispersion due to the viscosity, and found that it could be measurable in a number of materials, particularly ferromagnetic insulators. Other previous studies on lattice models~\cite{P_pol_Hvisc,DMRG_Hvisc,mengcheng} have calculated the Hall viscosity  using entanglement properties (momentum polarization), and have used the results to study the properties of the representative topological ground-state wave functions. 
Another recent study~\cite{Milovanovic} suggests that the Hall viscosity of a lattice Chern insulator may be related to a length scale associated with the Berry curvature at high symmetry points of the Brillouin zone. This is motivated by the idea that the low energy effective theory of Chern insulator is dual to an effective theory in reciprocal space subject to an effective magnetic field (Berry curvature). Interestingly, the end result of this last work matches the continuum regularized result mentioned above. 

Our twofold focus here is different from the previous work. We first address how strain deformations can be modeled in tight-binding lattice models where continuous translational symmetry does not exist, and, second, to what extent the lattice results coincide with the field theory calculations. In fact, our aim is not only to provide an alternative framework for numerical simulations, but also to relate the abstract notions in the field theory, such as geometric deformations, to the mechanical properties of the crystal, such as phonons, as was also done in Ref. \onlinecite{Maissam2012}. This is a step toward providing direct signatures of the Hall viscosity in experiments on TI materials. In addition, we present a 3D generalization of these approaches and use it to study the surface Hall viscosity of 3D TIs with magnetic surface layers.


To illustrate our methods we investigate three non-interacting models: (1) the Hofstadter model, i.e., the lattice version of the integer QHE, which primarily serves to benchmark our methods; (2) the Chern insulator lattice Dirac model, and (3) a 3D time-reversal invariant TI with magnetic layers on the surfaces. After the strain is implemented in the models, the Hall viscosity is computed using the Kubo formula~\cite{Kubo1966} in terms of the correlation function of stress operators $\hat{\cal T}_{ij}$ (see Ref.~\onlinecite{Read_Kubo} for a comprehensive discussion on this):
\begin{align} \label{eq:Kubo_visc}
\zeta_H &= \lim_{\omega\to 0} \frac{1}{\omega} \frac{1}{L^2} \int dt e^{i\omega t} \langle [\hat{\cal T}_{11}(t),\hat{\cal T}_{12}(0)] \rangle \nonumber \\
&=- \frac{2}{L^2}  \text{Im} \sum_{\substack{\nu\in occ. \\ \nu'\in unocc.}} \frac{\langle \nu|  \hat{\cal T}_{11}|\nu'\rangle \langle \nu' |\hat{ \cal T}_{12} |\nu \rangle}{(E_{\nu'}-E_\nu)^2},
\end{align}
where the system size is $L\times L$ and $|\nu\rangle$ denotes a single particle eigenstate of the Hamiltonian $H|\nu\rangle = E_\nu |\nu\rangle$.

Our calculations confirm that for the Hofstadter model in a weak (compared to the lattice scale) magnetic field, the lattice calculations coincide with the continuum expression. However,  we start to see deviations from the continuum limit as the magnetic field is increased and lattice effects become more important. In the Chern insulator model, we find that in the vicinity of the critical points where the topological phase transitions occur, the lattice results coincide with continuum limit predictions; however, away from the critical points, the value of the Hall viscosity is harder to determine and depends on our method of calculation. We also show that the momentum polarization approach~\cite{P_pol_Hvisc,DMRG_Hvisc} and our first method yield numerically identical results near the phase transition points.  For the 3D TI, we find that  the Hall viscosity at the surface of the 3D TI can be fit as a quadratic polynomial in the surface mass gap. The coefficient of the quadratic term also has an interesting dependence on the bulk gap. 
Our results for both the 2D Chern insulator and the gapped surface of a lattice Dirac model share the common feature that the Hall viscosity is continuous as the phase boundaries are crossed, and eventually asymptotes to zero deep in the trivial phase. The Hall viscosity due to the non-trivial 3D TI electronic structure adds an anomalous term to the surface phonon dynamics, the effects of which can in turn be experimentally observed. Our estimates show that these effects are well within the precision of current experimental technologies. We briefly discuss possible experimental scenarios where these effects can be investigated.

Our article is organized as follows: In Sec.~II, we present the two methods to describe the effects of strain in the tight-binding models. The three subsequent Secs.~III, IV, and V report the results for the three model Hamiltonians mentioned above,  possible experimental signatures of the Hall viscosity are discussed at the end of Sec.~V, and conclusions are outlined in Sec.~VI. There is also an appendix, which contains some extra details of derivations.

\section{Tight-binding models in the presence of strain}
Here, we introduce two different methods for modeling strain on a lattice in order to couple the electrons to variations in the background geometry, see Fig.~\ref{fig:lattice} for example.
We will be considering discrete lattice models represented by  a generic multi-orbital tight-binding model
\begin{align*}
\hat{H}= \sum_{r,r'}  c_r^\dagger t_{rr'} c_{r'},
\end{align*}
where $c_r^\dagger$ is a row vector of electron creation operators corresponding to the orbitals located in a unit cell at $r,$ and $t_{rr'}$ is an overlap matrix between two sites at $r$ and $r'$. 

\subsection{Minimal Coupling as a Gauge Field}
For our first method we show that the strain in a tight-binding model can be modeled as a generalized Peierls substitution.
Inspired by the analogy between the viscoelastic formalism and conventional Maxwell electromagnetism given in the field theoretical discussions of Refs.~\onlinecite{Taylor2011,Taylor2013,Hidaka2013}, we can look for a generalized minimal coupling in lattice models that must fulfill the following requirements:
\begin{enumerate}[i.]
\item 
The continuum limit must coincide with the field theory.
\item 
It must respect the $2\pi$ periodicity of the lattice momenta; in other words, the Brillouin zone needs to be well-defined during the deformation process.
\item
The \emph{distortion} field $w_{ij}$, to be defined below, must be a continuous variable.
\end{enumerate}
The third point is crucial as we are planning to study the variations of the Hamiltonian with respect to infinitesimal strain deformations and hence derive the stress operators.

Before we proceed, let us briefly review the aforementioned analogy in a more field-theoretical language. In Cartan's formulation of differential geometry in two spatial dimensions, a change in the metric is described by a set of local frame fields (vielbein) $\underline{e}_j(\textbf{x})$ where $j=1,2$. The $\mu$-th component of the frame field $\underline{e}_j(\textbf{x})$ is denoted by $\underline{e}^\mu_j(\textbf{x})$  (here $\mu=1,2$) in the local coordinate basis $(\partial_x, \partial_y)$ i.e.~$\underline{e}_j(\textbf{x})= \underline{e}^1_j(\textbf{x}) \partial_x +\underline{e}^2_j(\textbf{x}) \partial_y$. The metric tensor is defined in terms of the co-frame fields  $e^j(\textbf{x})=e^j_\mu(\textbf{x}) dx^\mu$ (dual to the frame fields, such that $e^i (\underline{e}_j)=\delta^i_j$) as  $g_{\mu\nu}= \delta_{ij} e^i_\mu e^j_\nu$, where $\delta^i_j=\delta_{ij}$ is the Kronecker delta. Hence, in the absence of spatial curvature the torsion tensor is determined by $T^j_{\mu\nu}= (de^j )_{\mu\nu}= \epsilon_{\mu\nu} \partial_\mu e^j_\nu$ where the spin connection has been chosen to be zero using the gauge-freedom in the absence of curvature. As discussed in Ref.~\onlinecite{Taylor2013}, the matter field is coupled to the strain through its momentum
\begin{align} \label{eq:p_gauge_cont}
p_j \to p_\mu \underline{e}^\mu_j = p_j - p_\mu w^\mu_j
\end{align}
where $\underline{e}^\mu_j= \delta^\mu_j- w^\mu_j,$ and $w^\mu_j= \partial_j u^\mu$ is the distortion (unsymmetrized strain) tensor. The shift in the momentum is reminiscent of the standard minimal coupling ($p_j - e A_j$) to the electromagnetic gauge field $A_j$; however, here the ``gauge field'' $w^\mu_j$ (overall four terms since $j=1,2$ and $\mu= 1,2$) is multiplied by the momentum. So, the ``charge" associated with this gauge field is indeed the momentum $p_\mu$. 

\begin{figure}
\centering
\includegraphics[scale=0.3]{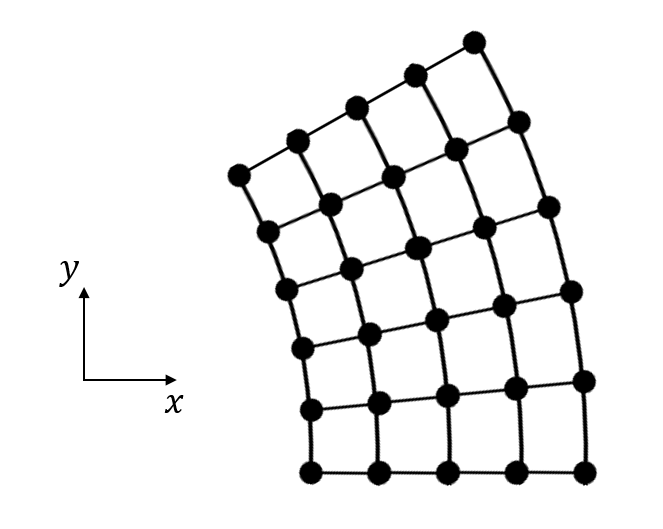}
\caption{\label{fig:lattice} Deformed lattice where the distortion tensor is given by $w_{22}=Gx$. Note that bonds along $x$ direction remain unchanged while the bonds in $y$ direction are stretched uniformly as a function of $x$.}
\end{figure}


Now consider the lattice Hamiltonian written in reciprocal space $\hat{H}= \sum_\textbf{k} c_\textbf{k}^\dagger h(\textbf{k}) c_\textbf{k}$. A uniform distortion, $w_{ij}$, (henceforth we only use lower indices for the distortion/strain fields in the lattice models, we will also be slightly imprecise and refer to a distortion $w_{ij}$ as a strain, and distinguish from the symmetrized version of the strain tensor by the symbol alone) is implemented as a shift in the lattice momenta:
\begin{align}  \label{eq:p_gauge}
k_i &\to k_i - w_{ji} \frac{\sin (k_ja)}{a} 
\end{align}
where $a$ is the lattice constant. This choice of coupling clearly satisfies all the requirements. In the continuum limit, $a\to 0$, the above expression is simplified to $k_i \to k_i - w_{ji} k_j$ which is consistent with Eq.~(\ref{eq:p_gauge_cont}). Note that there is a problem if one simply uses $k_i \to k_i - w_{ji} k_j$ as the minimal-coupling prescription for a lattice model because it does not satisfy the last two requirements simultaneously, i.e., the second condition is not met for arbitrary $w_{ji},$ and if we force it to meet this condition, then $w_{ji}$ has to be quantized and is no longer continuous (violation of the third condition).  Another remark in the definition of Eq.~(\ref{eq:p_gauge}) is that the conserved charge associated with this gauge symmetry is the discrete momentum operator\begin{align} \label{eq:disc_p}
\hat{p}_j = \frac{i}{2a} \sum_{\textbf{x}} (c_{\textbf{x}}^\dagger c_{\textbf{x}+\textbf{a}_j}- c_{\textbf{x}+\textbf{a}_j}^\dagger c_{\textbf{x}})
\end{align}
whose reciprocal-space representation is 
\begin{align*}
\hat{p}_j= \frac{1}{a} \sum_{\textbf{k}} \sin(k_j a)\ c^\dagger_{\textbf{k}}c_{\textbf{k}} \ .
\end{align*}

Now that we have specified the coupling to geometry let us describe the Hall viscosity response coefficient arising from this method. In order to compute the Hall viscosity as a susceptibility in the linear response formalism, we need to determine the generalized force (stress) operators
\begin{align*}
\hat{\cal T}^{(g)}_{ij} = \frac{\delta \hat{H}^{(g)}}{\delta w_{ij}}{\Big|}_{w=0}\ .
\end{align*}
We shall put the superscript $^{(g)}$ for all quantities used in connection to this generalized Peierls gauge-formalism. Using the transformation rule in Eq.~(\ref{eq:p_gauge}), one can easily relate the stress operators in reciprocal space, $\hat{\cal T}^{(g)}_{ij}=\sum_{\textbf{k}}  c^\dagger_{\textbf{k}} {\cal T}^{(g)}_{ij}(\textbf{k})\ c_{\textbf{k}} $, to the particle current operators ${\cal J}_{i}(\textbf{k}),$ and we find
\begin{align*}
{\cal T}^{(g)}_{11}(\textbf{k})&= \sin (k_1a)\ \frac{\delta h(\textbf{k})}{\delta k_1}{\Big|}_{w=0} = \sin (k_1a)\ {\cal J}_1(\textbf{k}), \\
{\cal T}^{(g)}_{12}(\textbf{k}) &= \sin (k_1a)\ \frac{\delta h(\textbf{k})}{\delta k_2}{\Big|}_{w=0} = \sin (k_1a)\ {\cal J}_2(\textbf{k}) ,
\end{align*}
which describe the flow of momentum. From the definition of the Hall viscosity, Eq.~(\ref{eq:Kubo_visc}), one can write
\begin{widetext}
\begin{align}
\zeta^{(g)}_H&=-\frac{2}{L^2} \text{Im} \sum_{\substack{\textbf{k}\\ \alpha\in \text{occ.} \\ \beta\in \text{unocc.}}}
\frac{\langle \alpha,\textbf{k} | {\cal T}^{(g)}_{11}(\textbf{k}) |\beta,\textbf{k}\rangle \langle  \beta,\textbf{k}|{\cal T}^{(g)}_{12}(\textbf{k}) |\alpha,\textbf{k} \rangle}{(E_{\alpha \textbf{k}}-E_{\beta \textbf{k}})^2}\nonumber \\
&=-\frac{2}{L^2} \text{Im} \sum_{\substack{\textbf{k}\\ \alpha\in \text{occ.} \\ \beta\in \text{unocc.}}}
\frac{\langle \alpha,\textbf{k} | \sin(k_1a) {\cal J}_{1}(\textbf{k}) |\beta,\textbf{k}\rangle \langle  \beta,\textbf{k}|\sin(k_1a) {\cal J}_{2}(\textbf{k}) |\alpha,\textbf{k} \rangle}{(E_{\alpha \textbf{k}}-E_{\beta \textbf{k}})^2} \nonumber \\
&=-\frac{1}{L^2} \text{Im} \sum_{\substack{\textbf{k}\\ \alpha\in \text{occ.} \\ \beta\in \text{unocc.}}}
 (\sin^2 (k_1a)+\sin^2 (k_2a)) \frac{\langle \alpha,\textbf{k} | {\cal J}_{1}(\textbf{k}) |\beta,\textbf{k}\rangle \langle  \beta,\textbf{k}|{\cal J}_{2}(\textbf{k}) |\alpha,\textbf{k} \rangle}{(E_{\alpha \textbf{k}}-E_{\beta \textbf{k}})^2} \nonumber \\
&= \frac{1}{2L^2} \sum_\textbf{k} (\sin^2 (k_1a)+\sin^2 (k_2a)) {\cal F}(\textbf{k}) \nonumber\\
&= \frac{1}{L^2} \sum_\textbf{k} {\cal B}^{(g)}(\textbf{k})
\end{align}
\end{widetext}
where $h(\textbf{k})|\alpha,\textbf{k}\rangle= E_{\alpha\textbf{k}}|\alpha,\textbf{k}\rangle$, $\alpha$ refers to the band index, and ${\cal B}^{(g)}(\textbf{k})$ is defined as the viscoelastic adiabatic curvature in the gauge coupling approach.
In the third line, we have used the assumed $C_4$ symmetry of the system (invariance under $k_1 \leftrightarrow k_2$) and write the expression in a symmetric fashion by including the contribution from $[\hat{\cal T}_{22},\hat{\cal T}_{21}]$. This assumption is only necessary to simplify our discussion and we could relax the rotation symmetry without much extra difficulty. The usual adiabatic curvature associated with $U(1)$ phase of the wave functions (Berry curvature) is denoted by ${\cal F}(\textbf{k})$:
\begin{align*}
{\cal F}(\textbf{k}) &=-2\ \text{Im} \sum_{\substack{\alpha\in \text{occ.} \\ \beta\in \text{unocc.}}}
 \frac{\langle \alpha,\textbf{k} | {\cal J}_{1}(\textbf{k}) |\beta,\textbf{k}\rangle \langle  \beta,\textbf{k}|{\cal J}_{2}(\textbf{k}) |\alpha,\textbf{k} \rangle}{(E_{\alpha \textbf{k}}-E_{\beta \textbf{k}})^2} .
\end{align*}

If we use the the gauge invariant formula of the Berry curvature we have
\begin{align} \label{eq:proj_visc_gauge}
\zeta^{(g)}_H=\frac{1}{2i L^2} \sum_\textbf{k} (\sin^2 k_1+\sin^2 k_2)  \epsilon_{ij} \text{tr}({\cal P}_\textbf{k} \partial_i {\cal P}_\textbf{k} \partial_j {\cal P}_\textbf{k})
\end{align}
where ${\cal P}_{\textbf{k}}= \sum_\alpha |\alpha,\textbf{k}\rangle \langle \alpha,\textbf{k}|$ is the projection operator onto the occupied states, and $\partial_i = \frac{\partial}{\partial k_i}$.  Using this relation will allow for a simple numerical computation of $\zeta_H^{(g)}$. 

Before we move on to the second method, it is useful to note that the existence of a conserved charge in this formalism provides two alternative formulas for the Hall viscosity besides the linear response (Kubo) formula: (a) the Streda formula and (b) momentum pumping/transport between the edges states using a Laughlin gauge argument. Let us discuss these in a bit more detail. 

One can write the Streda formula for the Hall viscosity in terms of bulk quantities following analogy to the electromagnetic response. Let us motivate the idea by first reviewing the electromagnetic response.
The Streda formula~\cite{Streda} for the Hall conductance is
\begin{align*}
\sigma_H = e \left( \frac{\partial M_z}{\partial \mu} \right)_{T,B} 
\end{align*}
where $e$ is the elementary charge,  $M_z$ is the bulk magnetization, $\mu$ is the chemical potential, $T$ is temperature, and $B$ denotes the magnetic field strength. The conjugate quantities to the magnetization and the chemical potential are magnetic field and particle number density, respectively. By means of thermodynamic conjugacy relations we obtain
\begin{align*}
\sigma_H = e \left( \frac{\partial n}{\partial B} \right)_{T,\mu} .
\end{align*}
Intuitively, this expression states that inserting a magnetic flux binds charge to that flux.
In a lattice model, one can look at the change in the particle density $n(x)= \sum_{\nu \in \text{occ.}} |\langle x|\nu(B)\rangle|^2$ as a function of the magnetic field ($B$) where the single particle states for a fixed $B$ are denoted by $|\nu(B)\rangle$. For small amplitudes of $B$ the density varies linearly with $B,$ and the Hall conductance can be read off from the slope.

Using this as a guide, we expect that applying a viscoelastic ``magnetic field'' would then add ``momentum charge'' into the system. The uniform viscoelastic ``magnetic field'' $G$ is determined by the viscoelastic ``vector potentials'' $w_{ij}$. For instance, the viscoelastic vector potential/distortion tensor  $w_{22}=G x$ (which is in a Landau gauge), corresponds to the lattice distortion shown in Fig.~\ref{fig:lattice}. For this choice of gauge, the system retains translational symmetry along the $y$-direction, and $k_2$ is a good quantum number.
Therefore, the following Streda formula~\cite{Hidaka2013} can be proposed for the Hall viscosity
\begin{align} \label{eq:Streda}
\zeta_{H}^{(g)}= \frac{\partial P_{2}}{\partial G}
\end{align}
where the transverse momentum ``charge'' density is found by $P_{2} (x)= \sum_{\nu \in \text{occ.}} \sin(k_2a)/a |\langle x|\nu(G)\rangle|^2$ ($x$ is in the bulk, far from edges) and $|\nu(G)\rangle$ is the eigenstate of the Hamiltonian subject to the viscoelastic magnetic field of strength $G$. 

\begin{figure}
\includegraphics[scale=0.65]{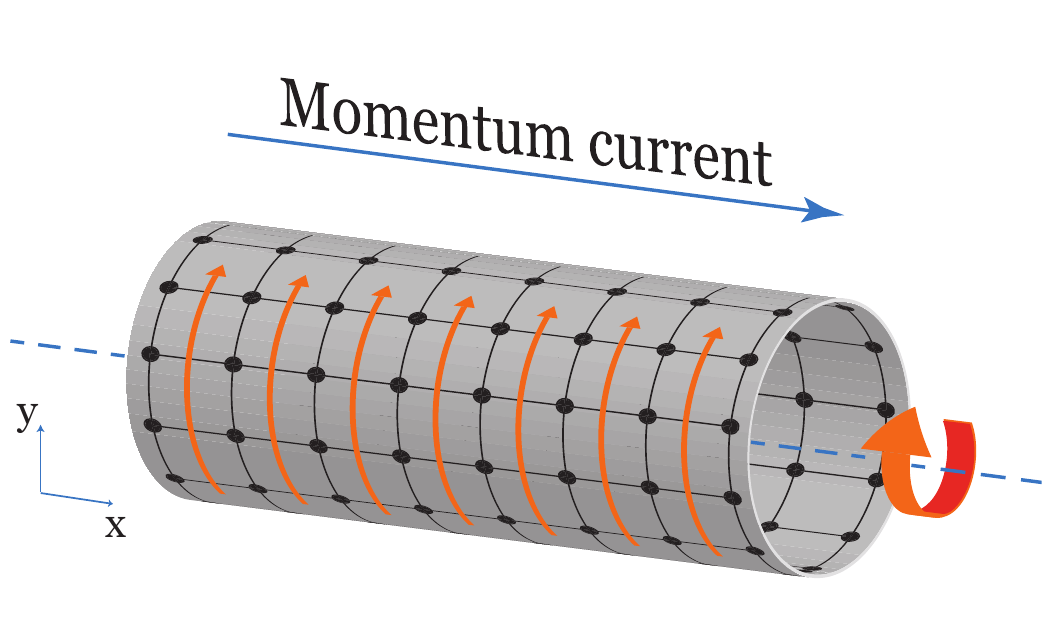}
\caption{\label{fig:spec_flow}  The generalized Laughlin experiment for the Hall viscosity. A viscoelastic flux is threaded through the cylinder, which one should think of as threading a dislocation through the cylinder. Electrons encircling the cylindrical hole will be translated by the Burgers' vector of the dislocation flux. For the case illustrated here the Burgers' vector is in the periodic $y$ direction. If the system has a Hall viscosity there will be a momentum current flowing in the $x$-direction carrying momentum pointing in the $y$-direction.}
\end{figure}

The second type of alternative viscosity calculation that can be used in the presence of a conserved charge relies on the momentum transport between opposite edge states when a viscoeleastic magnetic flux (dislocation) is threaded through the hole of a cylinder. To be explicit, let us consider edges realized in a cylindrical geometry (Fig.~\ref{fig:spec_flow}), where open boundary conditions are imposed along the $x$ direction, and $k_2$ is still a good quantum number in the periodic $y$ direction. A generalized Laughlin experiment~\cite{Laughlin_spec} can be done to measure the Hall viscosity. Threading a vicoelastic magnetic flux $\phi_G$ along the axis of cylinder changes the transverse momentum argument $k_2$ of the Hamiltonian $h(k_1,k_2)$ into $k_2-\phi_G \sin(k_2 a)/a$. Thus, as $\phi_G$ is increased, the spectrum of the edge states is modified and there is a net momentum transfer from one end to another\cite{Taylor2013}. The Hall viscosity is given by
\begin{align} \label{eq:Laughlin}
\zeta_H^{(g)} = \frac{1}{2} \frac{\partial \Delta P_2}{\partial \phi_G}
\end{align}in which $\Delta P_2= P_{2,R}-P_{2,L}$ is the difference in the momentum charge at the two edges of the cylinder. The momentum charge $P_{2,L(R)}$ at the left (right) edge is calculated by summing the momentum density over a few unit cells (greater than the penetration depth of the edge modes) at the left (right) edge.  The momentum density is given by  $P_2(x) = \sum_{\nu \in \text{occ.}} \sin(k_2 a) /a |\langle x|\nu(\phi_G) \rangle|^2$ where $|\nu(\phi_G)\rangle$ is the eigenstate of the Hamiltonian subject to the viscoelastic magnetic flux of $\phi_G$ (Fig.~\ref{fig:spec_flow}).

In Sec.~IV, we will do an explicit calculation using all three of these sub-methods: Kubo, Streda, and Laughlin and show that they all match for the Chern insulator model.

\subsection{Electron Geometric Coupling Through Lattice Distortions}

For the second computational formalism, the strain is modeled based on microscopic deformations of a tight-binding model. 
This type of approach for the electron-phonon coupling in tight-binding models has been carefully studied in graphene (see Ref.~\onlinecite{Castro2009} and references therein) and there are interesting proposed effects based on the predictions of these calculations. For example, it was shown that a non-uniform strain can lead to an effective magnetic field with opposite signs at two valleys~\cite{Guinea2010}, and this was subsequently experimentally  observed~\cite{Levy2010}. This type of method was first applied to the Hall viscosity problem by Ref.~\onlinecite{Maissam2012}. In that article they studied three models:
(1) the lowest Landau band in the Hofstadter model, although they ultimately only focus on the continuum limit; (2) a quantum spin Hall system in HgTe quantum wells\cite{BernevigScience,Molenkamp2007,PhysRevB.74.085308,PhysRevLett.101.146802,Yu2010}  in the presence of a small time-reversal breaking magnetization, where they investigate both the lattice and the continuum limit and discover that there is a discontinuity in the Hall viscosity at the transition to the  topological phase, and (3) a mean-field model of a $p_x+i p_y$ superconductor\cite{SrRu_RMP,ZhangRev}.
A common assumption in this approach, which we will also use, is that in modeling the strain, the lattice distortion is considered as an adiabatic process for electronic degrees of freedom. This implies that the phonon frequency must be much smaller than the electronic energy band gap. Moreover, we assume that the deformations are smooth  on lattice scales, which implies the phonon frequency to be smaller than the Debye frequency. Note that the latter assumption is not essential for lattice calculations and it is required only when one wants to take the continuum limit.

\begin{figure}
\begin{tikzpicture}
\node[anchor=center,inner sep=0] at (-0.5,0) {\includegraphics[scale=1]{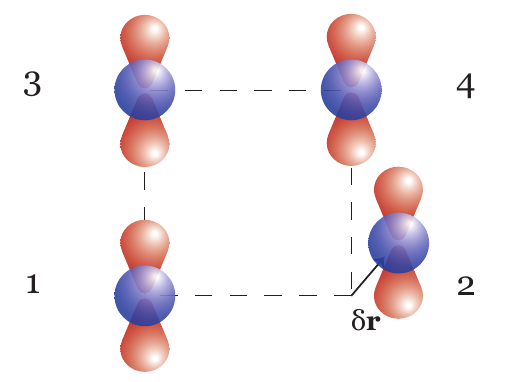}};
\draw[->] (-3.7,-1.8)--(-2.8,-1.8) node[below]{ $\textbf{a}_1$};
\draw[->] (-3.7,-1.8)--(-3.7,-1.1) node[left]{ $\textbf{a}_2$};
\end{tikzpicture}
\caption{\label{fig:hopp_strain} Illustration of the change in the overlap integrals as a result of strain.}
\end{figure}

In this subsection, we will derive a computational formula (Eq.~(\ref{eq:proj_visc_ph})) for the Hall viscosity based on this model for the applied strain and we will apply it to two 2D examples and one 3D example in subsequent sections:
(1) the Hofstadter model where we study both numerically and analytically the response of arbitrary integer filling fractions away from the continuum limit, and (2) the Chern insulator which is essentially half of the model considered for the quantum spin Hall effect in HgTe. In the latter case, however,  we find there is no discontinuity at the transition to the topological phase, which we discus further below. We should emphasize that this observation is also consistent with the field theory calculations. (3) We further obtain a 3D generalization of this formula and apply it to calculate the Hall viscosity at the surface of the 3D TI when time-reversal is broken on the surface.

Now let us introduce the method. A hopping matrix element $t(\textbf{r})$ in a tight-binding model is an overlap integral between two orbitals spatially separated by $\textbf{r}$. We can define the strain field to be a function of  the deviation $\delta\textbf{r}$ from the equilibrium value $\textbf{r}_0$. The linear order correction to the hopping matrix can be written as
\begin{align} \label{eq:eph_hop}
t(\textbf{r}_0+\delta\textbf{r}) \approx t(\textbf{r}_0)+ \frac{(\textbf{r}_0\cdot\delta\textbf{r})}{r_0} \frac{\partial t}{\partial r}{\Big|}_{\textbf{r}_0} + O(\delta r^2),
\end{align}
which is due to the change in the bond length (isotropic contribution) and exists for all orbitals. In addition to this term, if more than one orbital (or local degree of freedom) is present in each unit cell, there can be a correction to the hopping term between unlike orbitals due to an apparent rotation seen from  neighboring sites. For instance, Fig.~\ref{fig:hopp_strain} shows a lattice model with two types of orbitals, $s$ (blue) and $p_y$ (red), where the non-zero hopping terms along the $1-2$ bond $\textbf{a}_1$ are only between alike orbitals. As a result of the deformation $\delta \textbf{r}$, the hopping matrix between $s$ and $p_y$ orbitals becomes non-zero and proportional to the component of $\delta \textbf{r}$ perpendicular to the unperturbed lattice vector $\textbf{a}_1$,
\begin{align*}
t_{s,p_y}(\textbf{a}_1+\delta\textbf{r})\approx \frac{\textbf{n}\cdot(\textbf{a}_1\times\delta\textbf{r})}{|\textbf{a}_1|} t_{s,p_y}(\textbf{a}_2)
\end{align*}
where $\textbf{n}$ is the normal vector to the plane and $t_{s,p_y}(\textbf{a}_2)$ is the hopping amplitude between $s$ and $p_y$ orbitals in the vertical direction, $\textbf{a}_2$.
In general, $\delta\textbf{r}$ is related to the strain tensor $u_{ij}$ (phonon field). For the example in Fig.~\ref{fig:hopp_strain}, the hopping terms between sites 1 and 2 are modified as follows 
\begin{align*}
t_{s,s}(\textbf{a}_1+\delta\textbf{r}) &\approx t_{s,s}(\textbf{a}_1) - u_{11}\ t_{s,s}' \nonumber \\
t_{p_y,p_y}(\textbf{a}_1+\delta\textbf{r}) &\approx t_{p_y,p_y}(\textbf{a}_1) - u_{11}\ t_{p_y,p_y}' \nonumber \\
t_{s,p_y}(\textbf{a}_1+\delta\textbf{r}) &\approx  u_{12}\ t_{s,p_y}(\textbf{a}_2)
\end{align*}
where  $t_{\ell,\ell}'= -a \partial t_{\ell,\ell}/\partial r|_{a} $ is the derivative with respect to the lattice constant which is originally $|\textbf{a}_1|=|\textbf{a}_2|=a$. We remind the reader that we will assume $C_4$ symmetry for simplicity, and the overlap integrals could be modified in a more complicated manner if the lattice vectors are not orthogonal. 

The perturbed Hamiltonian can then be written as a function of $u_{ij}$ and we define the strain operators by
\begin{align*}
\hat{\cal{T}}^{(p)}_{ij} = \frac{\delta \hat{H}^{(p)}}{\delta u_{ij}}{\Big|}_{u=0}
\end{align*}
on which the superscript $^{(p)}$ is placed for all quantities in this formalism.
Following the linear response formula for the Hall viscosity, we start with
\begin{widetext}
\begin{align}
\zeta^{(p)}_H=&- \frac{2}{L^2} \text{Im} \sum_{\substack{\textbf{k}\\ \alpha\in \text{occ.} \\ \beta\in \text{unocc.}}}
\frac{\langle t,\alpha,\textbf{k} |{\cal T}_{11}(\textbf{k}) |t,\beta,\textbf{k}\rangle \langle t, \beta,\textbf{k}|{\cal T}_{12}(\textbf{k}) |t,\alpha,\textbf{k} \rangle}{(E_{\alpha \textbf{k}}(t)-E_{\beta \textbf{k}}(t))^2}{\Big|}_{t_0} \nonumber \\
=&-\frac{2}{L^2} \text{Im} \sum_{\substack{\textbf{k}\\ \alpha\in \text{occ.} \\ \beta\in \text{unocc.}}}
\frac{\langle t,\alpha,\textbf{k} | [\partial_{u_{11}},{h}_t(\textbf{k})] |t,\beta,\textbf{k}\rangle \langle t,\beta,\textbf{k}|[\partial_{u_{12}},{h}_t(\textbf{k})] |t,\alpha,\textbf{k} \rangle}{(E_{\alpha \textbf{k}}(t)-E_{\beta \textbf{k}}(t))^2}{\Big|}_{t_0} \nonumber \\
=&-\frac{2}{L^2} \text{Im} \sum_{\substack{\textbf{k}\\ \alpha\in \text{occ.} \\ \beta\in \text{unocc.}}}
\frac{\langle t,\alpha,\textbf{k} |(\partial_{u_{11}}{h}_t(\textbf{k})-{h}_t(\textbf{k}) \partial_{u_{11}})|t,\beta,\textbf{k}\rangle \langle t,\beta,\textbf{k}|(\partial_{u_{12}}{h}_t(\textbf{k})-{h}_t(\textbf{k}) \partial_{u_{12}})|t,\alpha,\textbf{k} \rangle}{(E_{\alpha \textbf{k}}(t)-E_{\beta \textbf{k}}(t))^2}{\Big|}_{t_0} \nonumber \\
=&-\frac{2}{L^2} \text{Im} \sum_{\substack{\textbf{k}\\ \alpha\in \text{occ.} \\ \beta\in \text{unocc.}}}
\frac{\langle t,\alpha,\textbf{k} |(E_{\alpha\textbf{k}}-E_{\beta\textbf{k}})\partial_{u_{11}})|t,\beta,\textbf{k}\rangle \langle t,\beta,\textbf{k}|(E_{\alpha\textbf{k}}-E_{\beta \textbf{k}})\partial_{u_{12}}|t,\alpha,\textbf{k} \rangle}{(E_{\alpha \textbf{k}}(t)-E_{\beta \textbf{k}}(t))^2}{\Big|}_{t_0} \nonumber \\
=&-\frac{2}{L^2} \text{Im} \sum_{\substack{\textbf{k}\\ \alpha\in \text{occ.} \\ \beta\in \text{unocc.}}}
\langle t,\alpha,\textbf{k} |\partial_{u_{11}}|t,\beta,\textbf{k}\rangle \langle t,\beta,\textbf{k}|\partial_{u_{12}}|t,\alpha,\textbf{k} \rangle {\Big|}_{t_0}  \nonumber \\
=& \frac{1}{L^2} \sum_{\textbf{k}}{\cal B}^{(p)}(\textbf{k}) \\ 
\nonumber
\end{align}
\end{widetext}
where $h_t(\textbf{k}) |t,\alpha,\textbf{k}\rangle= E_{\alpha \textbf{k}}(t) |t,\alpha,\textbf{k}\rangle$ is the eigenstate of the Hamiltonian, $h_t(\textbf{k})$, with the set of hopping matrix elements collectively denoted by $t$. $t_0$ labels the original (unperturbed) values of hopping amplitudes. The \emph{viscoelastic} adiabatic curvature ${\cal B}^{(p)}(\textbf{k})$ over the occupied states is also introduced.   
 The above expression can be recast in a gauge invariant form
\begin{align} \label{eq:proj_visc_ph}
\zeta^{(p)}_H=-\frac{2}{L^2} \text{Im} \sum_\textbf{k} \text{tr}({\cal P}_\textbf{k} (\partial_{u_{11}} {\cal P}_\textbf{k}) (\partial_{u_{12}} {\cal P}_\textbf{k}))\ .
\end{align}
Note that the dependence on $u_{ij}$ comes from the dependence of the hopping matrix elements on the strain. We also urge the reader not to confuse the parameter dependence on the hopping amplitudes that we have denoted by $t,$ and the time-coordinate, which does not enter any of the previous expressions.

We have now completed the introduction of our methods and in the following sections we will investigate several lattice models with continuum limits that have been a subject of great interest. The  first model we look at is the Hofstadter model, which leads to the standard integer QHE Landau level problem in the continuum limit. The second and  third models are described by massive Dirac Hamiltonians in the long-wavelength limit, i.e., the minimal models for TIs. We compare the lattice results in each case with their continuum limit counterparts, and discuss the agreement between the two limits  and why, or why not, we should have such an expectation.

\section{Example 1: The Hofstadter Model}

\begin{figure}
\centering
\includegraphics[scale=0.72]{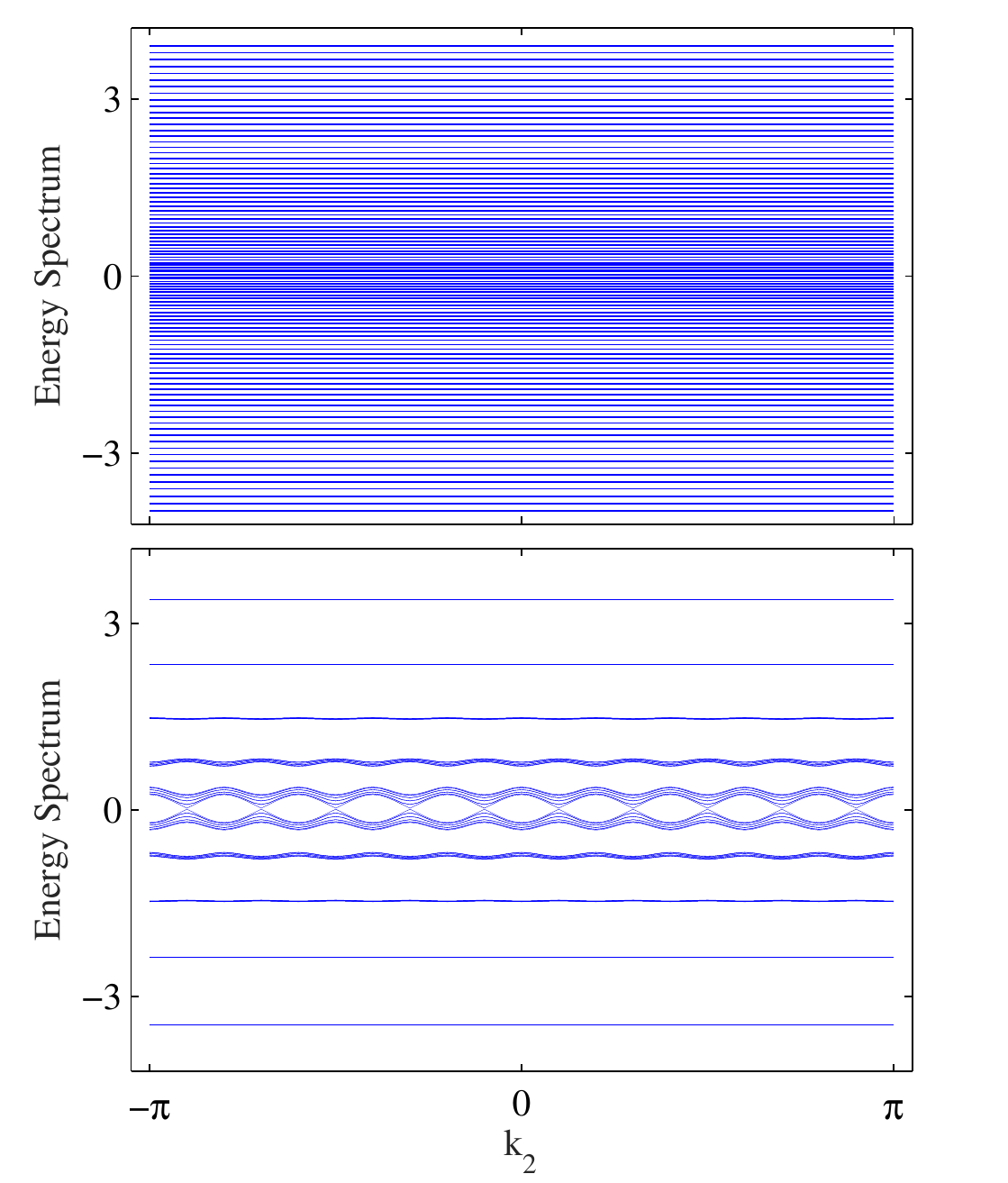}
\caption{\label{fig:LL_spec} The spectrum of the Hofstadter model for two values of the magnetic field $\phi/\phi_0=1/20$ (top) and $1/10$ (bottom). Parameters are $t=100\tilde{t}=2$, $a=1$ and the system size is $100\times 100$. }
\end{figure}

\begin{figure}
\centering
\includegraphics[scale=.72]{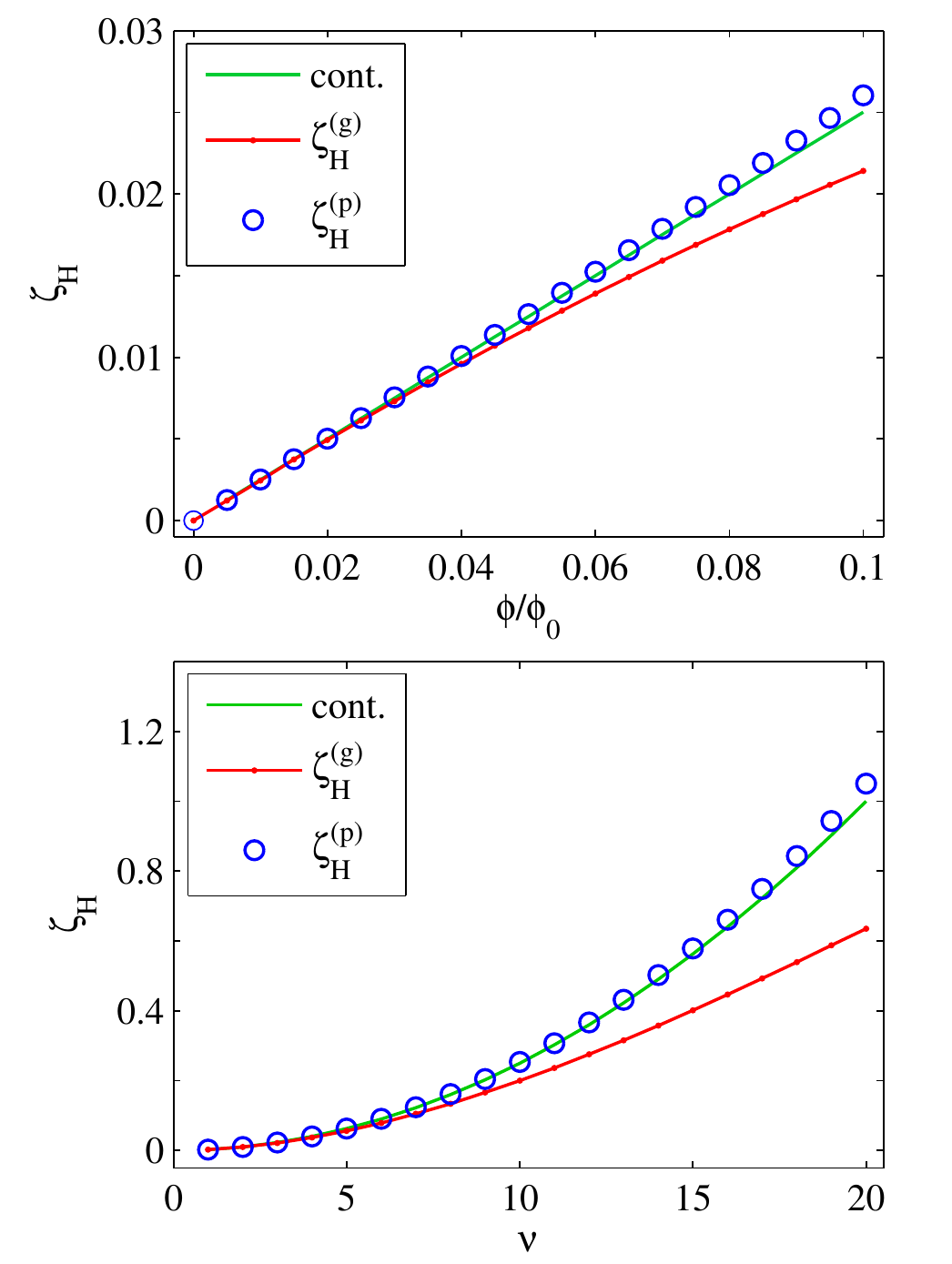}
\caption{\label{fig:Hoffs}  Top: the Hall viscosity at $\nu=1$ IQHE as a function of the magnetic flux per plaquette $\frac{\phi}{\phi_0}$. Bottom: the Hall viscosity vs the filling fraction $\nu$ (integer fillings) where $\frac{\phi}{\phi_0}=0.01$ is kept fixed. $\zeta_H$ is in units of $\hbar/2\pi a^2$ where $a$ is the lattice constant.}
\end{figure}

Consider the Hamiltonian of a single band tight-binding model on a square lattice subject to a magnetic field~\cite{Hofs}
\begin{align}\label{eq:singband}
\hat{H} = -\frac{1}{2} \sum_{\langle i,j \rangle} t e^{i A_{ij}} c_i^\dagger c_j - \frac{1}{2} \sum_{\langle\langle i,j \rangle\rangle} \tilde{t} e^{i A_{ij}} c_i^\dagger c_j
\end{align}
where ${\scriptstyle \langle\ \rangle}$ and ${\scriptstyle  \langle\langle\ \rangle\rangle}$ refer to the nearest neighbor and next nearest neighbor sites respectively, and
\begin{align*}
A_{ij}= \int_i^j \textbf{A}(\textbf{x})\cdot d\textbf{x}
\end{align*}
with the choice of Landau gauge ${\bf A}=B (0,x)$. The magnetic field is $B=\phi/a^2$ where the flux per plaquette is denoted by $\phi=p/q$ ($p$ and $q$ are coprime integers) in units of $\phi_0=h/e$.
 We define a supercell of size $q\times 1$ and derive the Hamiltonian in reciprocal space:
\begin{align*}
\hat{H} = \sum_\textbf{k} c_\textbf{k}^\dagger h(\textbf{k}) c_\textbf{k}
\end{align*}
where $c^\dagger_\textbf{k}= (c^\dagger_{1,\textbf{k}},\dots,c^\dagger_{q,\textbf{k}}),$ and the first index labels the sublattice position within the magnetic supercell. The Hamiltonian matrix h(\textbf{k}) reads

\begin{align*}
 \left( \begin{array}{ccccc}
\Delta_1(\textbf{k}) & \Xi_1(\textbf{k}) & & & \Xi_q(\textbf{k}) e^{ik_1a} \\
\Xi_1(\textbf{k}) &\Delta_2(\textbf{k}) & \Xi_2(\textbf{k}) & & \\
 & \Xi_2(\textbf{k}) &\cdot & \cdot & \\
 &  &\cdot & \cdot & \Xi_{q-1}(\textbf{k})\\
 \Xi_q(\textbf{k}) e^{-ik_1a}  &  &  & \Xi_{q-1}(\textbf{k}) & \Delta_q(\textbf{k})
\end{array} \right)
\end{align*}
in which $\Delta_x(\textbf{k})= -t \cos(k_2a-Bax)$, and $\Xi_x(\textbf{k})=-t/2 -\tilde{t} \cos(k_2a-Ba x-\phi/2)$. The spectrum of this Hamiltonian is given by the Landau bands (see Fig.~\ref{fig:LL_spec}). We can then use Eqs.~(\ref{eq:proj_visc_gauge}) and (\ref{eq:proj_visc_ph})  to calculate the Hall viscosity. The results are plotted in Fig.~\ref{fig:Hoffs}. In the lowest Landau level, Fig.~\ref{fig:Hoffs}(top), there is a remarkable agreement with the continuum expression up to a field strength of $\phi/\phi_0=3/50$. After that, the effects of the lattice become prevalent and we find a deviation from the continuum expression. In Fig.~\ref{fig:Hoffs}(bottom), we compute $\zeta_H$ for higher integer filling fractions and observe that the lattice calculations coincide with the continuum results up to $\nu=10$. In both plots, the electron-phonon coupling method seems to stay closer to the continuum results over a wider range than the gauge coupling method. Since, \emph{a priori}, we do not know what the correct value for the Hall viscosity is in a lattice system, we cannot identify which method, if either, is correct, only that they both reproduce the continuum limit, and both deviate when lattice effects become important.

Let us now show analytically that the continuum limits of both formulas lead to the same results. 
We will derive the continuum limit of the Hamiltonian in the absence of $B,$ and then add $B$ back in afterward.
In the gauge coupling formalism, we expand the Hamiltonian around $\textbf{k}=(0,0)$ as
\begin{align*}
h^{(g)}_{B=0}(\textbf{k})
\simeq& (\frac{t}{2}+ \tilde{t})[ (k_1-w_{11} k_1-w_{21} k_2)^2 \nonumber \\
&\  \ \ \ \ \ \ \ \ \ +(k_2-w_{22} k_2-w_{12} k_1)^2] +\dots \nonumber \\
=&  \frac{1}{2m} k_i g^{(g)}_{ij} k_j
\end{align*}
where the mass is $m=(t/2+\tilde{t})^{-1}/2,$ and the metric is given by
\begin{align}\label{eq:IQHE_g_metric}
g^{(g)}= \left( \begin{array}{cc}
1-2w_{11} & -2w_{12} \\
-2w_{21} & 1-2w_{22}
\end{array} \right)\ .
\end{align}
In the presence of the magnetic field, $k_i$ is simply replaced by $k_i-A_i$ (setting the charge $e=1$). The stress operators are found to be
\begin{align*}
{\cal T}^{(g)}_{11}&= \frac{{D}_1^2}{m} ,  \\
{\cal T}^{(g)}_{12}&=  \frac{1}{2m}   ({D}_1 {D}_2+{D}_2 {D}_1) ,
\end{align*}
where ${D}_i=-i \frac{\partial}{\partial x_i} - A_i$ is the electromagnetic covariant derivative. We introduce the ladder operators using the fact that $[{D}_2,{D}_1]=i B$
\begin{align} 
a &= \sqrt{\frac{1}{2B}} (D_2+iD_1),  \nonumber \\
\label{eq:LL_cr}
a^\dagger &= \sqrt{\frac{1}{2B}} (D_2-iD_1) ,
\end{align}
and the strain operators become
\begin{subequations}
\begin{align}
{\cal T}^{(g)}_{11}&= \frac{B}{2m} (a+a^\dagger)^2 ,  \\
{\cal T}^{(g)}_{12}&=  i \frac{B}{2m} (a^2-a^{\dagger2}) \ .
\end{align}
\end{subequations}
Let us compute the response for the $n$-th Landau level 
\begin{align*}
\zeta^{(g)}_H(n)=& -\frac{2}{2\pi \ell_B^2} \ \text{Im} \sum_{n'} \frac{\langle n|  {\cal T}^{(g)}_{11}|n'\rangle \langle n' |{\cal T}^{(g)}_{12} |n \rangle}{(E_n'-E_n)^2} \nonumber \\
=& -\frac{B^2}{4\pi m^2 \ell_B^2}    \sum_{n'}\frac{\langle n| (a+a^\dagger)^2  |n'\rangle \langle n' |a^2-a^{\dagger2} |n \rangle}{(E_n'-E_n)^2} \nonumber \\
=& \left(\frac{2n+1}{4}\right) \frac{1}{2\pi \ell_B^2} .
\end{align*}
Here $|n\rangle$ is the $n$-th Landau eigenstate $a^\dagger a|n\rangle = n |n\rangle,$ and the factor of $1/L^2$ was  canceled by the degeneracy of the Landau level $L^2/2\pi \ell_B^2$. Hence, for integer fillings  $\nu\geq 1$, we can use the above results to derive
\begin{align} \label{eq:IQHE_HV}
\zeta_H^{(g)}= \sum_{n=0}^{\nu-1} \zeta^{(g)}_H(n)= \frac{\hbar}{8\pi \ell_B^2} \nu^2
\end{align}
which is the same as the original findings of Avron \emph{et al.} (and also L\'{e}vay) ~\cite{Avron1995,Levay1995,Avron_multi}  for the integer QHE viscosity response under modular deformations in the metric.

Let us now proceed to the second method. The electron-phonon coupling formalism is modeled by the following changes in the hopping amplitudes
\begin{align*}
 t_{\textbf{x},\textbf{x}+\textbf{a}_1} &\to t - t' u_{11}, \nonumber  \\
 t_{\textbf{x},\textbf{x}+\textbf{a}_2} &\to t-  t' u_{22} , \nonumber \\
\tilde{t}_{\textbf{x},\textbf{x}+\textbf{a}_1\pm\textbf{a}_2} &\to \tilde{t}- \tilde{t}'( \pm u_{12} +\frac{1}{2} (u_{11} + u_{22})   )
\end{align*}
in which $t'=- a \frac{\partial t}{\partial r}|_{r=a}$, $\tilde{t}'=- \sqrt{2} a\frac{\partial \tilde{t}}{\partial r}|_{r=\sqrt{2}a},$ and we have used the identity $u_{12}=u_{21}$. Let us now look at the continuum limit in this formulation
\begin{align*}
h^{(p)}_{B=0}(\textbf{k})\simeq& (\frac{t}{2}+ \tilde{t}) (k_1^2+k_2^2)  - \frac{t'}{2} (u_{11} k_1^2+u_{22} k_2^2) \nonumber \\ &- \frac{\tilde{t}'}{2} (u_{11}+u_{22})( k_1^2+ k_2^2) -2\tilde{t}' u_{12} k_1 k_2 \nonumber \\
=&  \frac{1}{2m} k_i g^{(p)}_{ij} k_j
\end{align*}
where the mass is defined as $m=(t/2+\tilde{t})^{-1}/2$ and the metric is 
\begin{align} \label{eq:IQHE_eph_metric}
g^{(p)}=\mathbbm{1}- \left( \begin{array}{cc}
\alpha_{1} u_{11}+ \alpha_2 u_{22} & 2\alpha_2 u_{12} \\
2\alpha_2 u_{12} &  \alpha_{1}u_{22}+ \alpha_2 u_{11}
\end{array} \right),
\end{align}
where $\alpha_{1}=m(t'+\tilde{t}')$, and $\alpha_2=m\tilde{t}'$.
Notice that the phonon field $u_{ij}$, as it appears in the metric, is multiplied by the electron-phonon coupling coefficients $\alpha_{i}$.  We will see below that this will give a pre-factor in the final expression for the Hall viscosity.
 
Again, the effect of the magnetic field is recovered by substituting $k_i$ with $ k_i - A_i$. The stress operators can be found easily
\begin{align*}
{\cal T}^{(p)}_{11}&= \frac{1}{2} ((t'+\tilde{t}') {D}_1^2 +\tilde{t}'  {D}_2^2) , \\
{\cal T}^{(p)}_{12}&=  \tilde{t}' ({D}_1{D}_2+{D}_2{D}_1) ,
\end{align*}
which can be rewritten in terms of the ladder operators (Eq.~(\ref{eq:LL_cr}))
\begin{align*}
{\cal T}^{(p)}_{11}&= \frac{B}{4} (t' (a+a^\dagger)^2 +\tilde{t}' (aa^\dagger+a^\dagger a)) , \\
{\cal T}^{(p)}_{12}&=  i B\tilde{t}' (a^2-a^{\dagger 2}) .
\end{align*}
The Kubo formula for the $n$-th Landau level yields 
\begin{align*}
\zeta^{(p)}_H(n) =& -\frac{2}{2\pi \ell_B^2}\ \text{Im} \sum_{n'} \frac{\langle n|  {\cal T}^{(p)}_{11}|n'\rangle \langle n' |{\cal T}_{12}^{(p)} |n \rangle}{(E_n'-E_n)^2}  \nonumber \\
=&-\frac{1}{2\pi \ell_B^2} \frac{\tilde{t}'t'B^2}{2}  \sum_{n'} \frac{n(n-1)\delta_{n',n-2}}{(2\frac{B}{m})^2} \nonumber\\
&\ \ \ \  \ \  \ \ \ \ \ \ \ \  \ \ \ \ \  \  \  \ \ \ \ -\frac{(n+1)(n+2)\delta_{n',n+2}}{(2\frac{B}{m})^2} \nonumber \\
\label{eq:LL_ff_phonon}
=& \tilde{t}'t'm^2 \frac{(2n+1)}{4}  \frac{1}{2\pi \ell_B^2}.
\end{align*}
For filling fraction $\nu$, the total Hall viscosity is found to be
\begin{align}
\zeta_H^{(p)}= \frac{4\tilde{t}'t'}{(\frac{t}{2}+\tilde{t})^2} \frac{\hbar}{8\pi \ell_B^2}  \sum_{n=0}^{\nu-1} (2n+1)=  \frac{4\tilde{t}'t'}{(\frac{t}{2}+\tilde{t})^2} \frac{\hbar}{8\pi \ell_B^2} \nu^2 
\end{align}
where the usual Hall viscosity $(2n+1)/8\pi \ell_B^2$ is multiplied by a non-universal dimensionless pre-factor which depends on the electron-phonon coupling constants $t'$, and $\tilde{t}'$ and the band parameters. This does not contradict Eq.~(\ref{eq:IQHE_HV}) since the phonon-induced changes in the underlying geometry of the electrons  in Eq.~(\ref{eq:IQHE_eph_metric}) have extra coefficients depending on the electron-phonon coupling constants (as opposed to Eq.~(\ref{eq:IQHE_g_metric})), and hence the phonon Hall viscosity is proportional (and  not equal) to the usual (gravitational) Hall viscosity. 
Note that above, when we showed the electron-phonon results in Fig. \ref{fig:Hoffs}, we divided the data obtained from Eq.~(\ref{eq:proj_visc_ph}) by this pre-factor to compare only the magnetic field dependent factor. For the other models below we will not need to make this adjustment. 

We also note that the lattice effects on the Hall viscosity in the Hofstadter model has been studied recently in Ref.~\onlinecite{Thomas} using methods based on the momentum transport and momentum (entanglement) polarization. The Hall viscosity of Dirac electrons subject to a magnetic field has also been calculated~\cite{Thomas,Kimura}.
In addition, a semi-classical derivation~\cite{Rudro} for the Hall viscosity, which includes an analysis of the Hofstadter model, has been worked out. It would be interesting, in future work, to carefully compare each approach.


\section{Example 2: Chern Insulator Model}

For our next example we consider a (2+1)D lattice Dirac model for the Chern Insulator (CI) on a square lattice~\cite{Haldane_CI,Bernevig2013}:
\begin{align} \label{eq:CI_model}
\hat{H}=& \frac{1}{2} \sum_{\substack{\textbf{x},s}} {\Big[} c_{\textbf{x}+\textbf{a}_s}^\dagger (i t_s \sigma_s - r \sigma_3) c_\textbf{x} +\text{h.c.} {\Big]} + m \sum_\textbf{x} c_\textbf{x}^\dagger \sigma_3 c_\textbf{x}
\end{align}
where $\textbf{a}_s\  (s=1,2)$ are the square lattice basis vectors, $c^\dagger_\textbf{x}=(c^\dagger_{1,\textbf{x}},c^\dagger_{2,\textbf{x}})$ is the electron creation operator with two orbital (or spin) indices, and ($\sigma_1$, $\sigma_2$, $\sigma_3$) are Pauli matrices. The parameter $r>0$ is the hopping amplitude between identical orbitals, and $t_1$ and $t_2$ denote the hopping amplitudes between opposite orbitals (we choose the $C_4$ symmetric case $t_1=t_2=t>0$).

As the mass parameter $m$ is varied, we obtain the following phases:
\begin{enumerate}[(a)]
\item Topological Chern insulator phases: $-2r<m<0$ with $C=-1$ and $0<m<2r$ with $C=1$. At the critical point between these phases ($m=0$), there are two Dirac nodes at $(0,\pi)$ and $(\pi,0)$ in the Brillouin zone.
\item Trivial phases: $|m|>2r$. The critical theory at the critical points between the trivial and topological phases are described by a single Dirac node at $(0,0)$ or $(\pi,\pi)$ for $m=2r$ and $m=-2r$ respectively.
\end{enumerate}
For this model we calculate the Hall viscosity using both methods as $m$ is varied. We  show that the calculated viscosities have the following generic properties:
\begin{enumerate}[i.]
\item 
$\zeta_H$ vanishes at the critical point $m=0$ where the Chern number changes from $1$ to $-1$.
\item The Hall viscosity is continuous throughout the phase diagram.
\item
The Hall viscosity is generically finite in the topological phases \emph{and} in the trivial phases, but becomes zero deep in the trivial phases ($m\to \pm \infty$). The rate at which $\zeta_H$ goes to zero in the trivial phase is determined by the hopping amplitude $t$ (which characterizes the gapless Dirac Fermi-velocity).
\end{enumerate}
All three properties are consistent with the field theory considerations in Refs.~\onlinecite{Taylor2011,Taylor2013}.

Let us now discuss the details of the calculations. Using the gauge coupling formalism, we write the deformed Hamiltonian in  reciprocal space as
\begin{align} \label{eq:Chern_gauge}
\hat{H}^{(g)} =& \sum_{\textbf{k}}  c_{\textbf{k}}^\dagger {\Big[}i t \sigma_2\ \sin(k_2-w_{22} {p}_2-w_{12} {p}_1) \nonumber \\
&\ \ \ \  - r \sigma_3 \cos(k_2-w_{22} {p}_2-w_{12} {p}_1) {\Big]} c_\textbf{k}  \nonumber \\
&+ \sum_{\textbf{k}}  c_{\textbf{k}}^\dagger  {\Big[}i t \sigma_1\ \sin(k_1-w_{11} {p}_1-w_{21} {p}_2) \nonumber \\
&\ \ \ \ - r \sigma_3 \cos(k_1-w_{11} {p}_1-w_{21} {p}_2) {\Big]} c_\textbf{k}  \nonumber \\
&+ m \sum_\textbf{k} c_\textbf{k}^\dagger \sigma_3 c_\textbf{k}
\end{align}
where $p_i=\sin k_i$ is the discrete momentum defined in Eq.~(\ref{eq:disc_p}). 
This expression immediately gives the stress-current operators  $\hat{\cal{T}}^{(g)}_{ij} = \sum_{\textbf{k}} c^\dagger_\textbf{k} {\cal T}^{(g)}_{ij} (\textbf{k}) c_\textbf{k}$ where
\begin{align*}
{\cal T}^{(g)}_{11} (\textbf{k})&= (\sigma_1 t\cos k_1a + \sigma_3 r\sin k_1a) \sin k_1a, \\
{\cal T}^{(g)}_{12} (\textbf{k})&= (\sigma_2 t\cos k_2a + \sigma_3 r\sin k_2a) \sin k_1a.
\end{align*}
The lattice constant is denoted by $a,$ and the total number of sites is $N=L^2/a^2$. So, the Hall viscosity  is given
 by (see Appendix A)
\begin{widetext}
\begin{align} \label{eq:CI_gauge}
\zeta^{(g)}_H= \frac{1}{L^2} \sum_{\textbf{k}} {\cal B}^{(g)} (\textbf{k}) = \frac{1}{4Na^2}  \sum_{\textbf{k}} \frac{t^2 (\sin^2 k_1a + \sin^2k_2a) (m\cos k_1a \cos k_2a - r (\cos k_1a+\cos k_2a ) )}{\left[ (m - r\cos k_1a - r \cos k_2a)^2 + t^2 (\sin^2 k_1a + \sin^2k_2a)\right]^{3/2}}.
\end{align}
\end{widetext}

\begin{figure}
\includegraphics[scale=0.8]{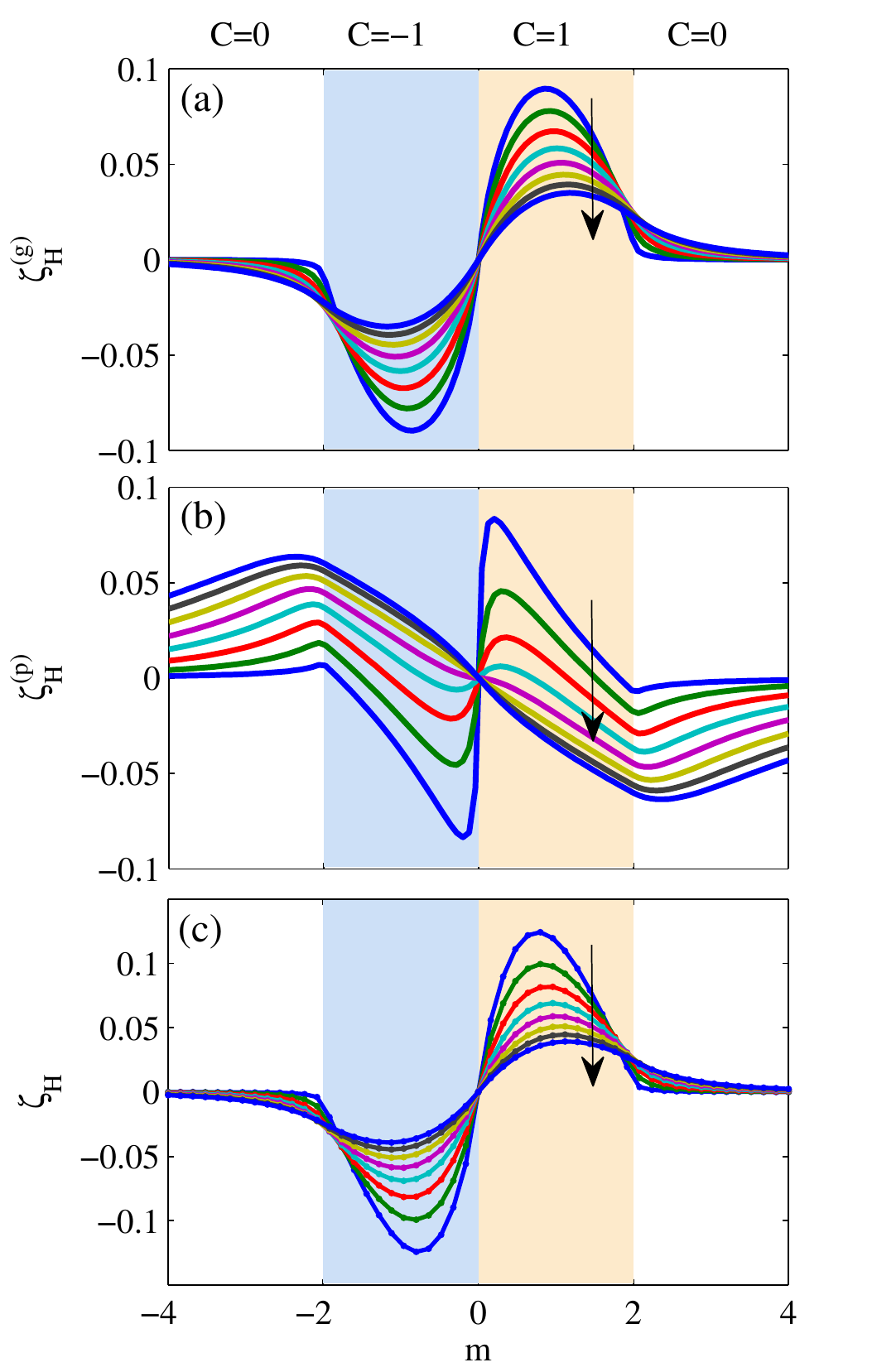}
\caption{\label{fig:CI}  The Hall viscosity (in units of $\hbar/a^2$) as a function of mass for various values of hopping amplitude $t$ in the range $[0.25,2]$. The graphs are calculated from the Kubo formula in the thermodynamic limit $N\to \infty$, (a) Eq.~(\ref{eq:CI_gauge}), (b) Eq.~(\ref{eq:CI_phonon}), and (c) momentum polarization method~\cite{P_pol_Hvisc,DMRG_Hvisc}. The arrow indicates the direction in which $t$ increases with steps of $0.25$. The colored regions show the topological phases and the Chern number ($C$) is indicated at the very top. The other hopping term is fixed $r=1$.}
\end{figure}

\begin{figure}
\subfigure{\includegraphics[scale=0.8]{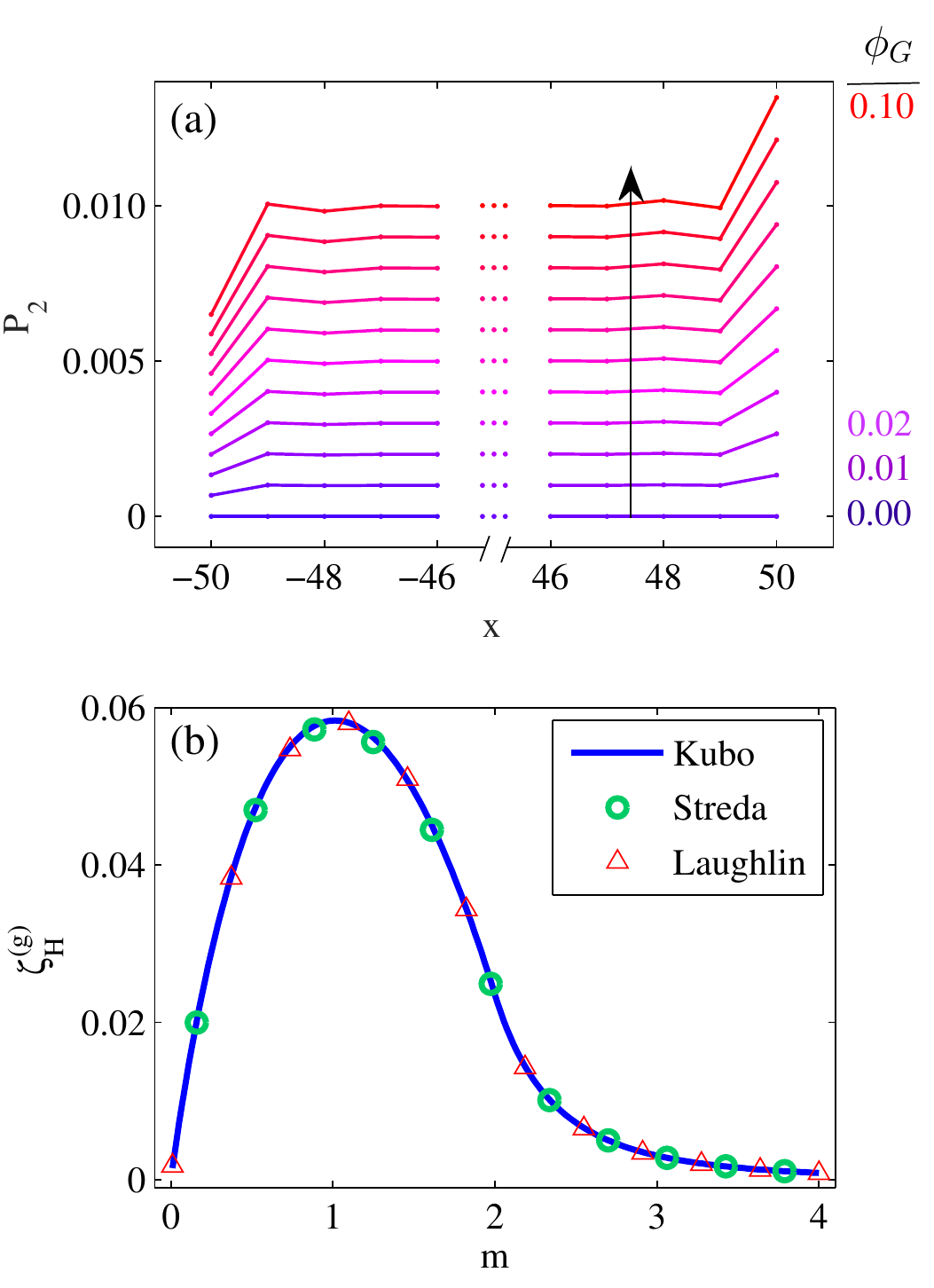}}
\caption{\label{fig:CI_comp}  (a) The illustration of the generalized Laughlin viscoelastic gauge experiment (see Fig.~\ref{fig:spec_flow}). Each curve shows the momentum density profile (in units of $\hbar/a$) along the cylinder.  The shift in the vertical direction is for presentation. Along the arrow from bottom to top the viscoelastic magnetic flux $\phi_G$ is increased from $\phi_G=0$ to $\phi_G=0.1$ with steps of $0.01$ . As a result, the momentum charge accumulated at the edges gradually increases. The bulk gap is fixed $m=1$. (b) A comparison of the Hall viscosity (in units of $\hbar/a^2$) in the gauge coupling formalism using the Kubo formula Eq.~(\ref{eq:CI_gauge}), the Streda formula, Eq.~(\ref{eq:Streda}), and the spectral flow of the momentum at the edges, Eq.~(\ref{eq:Laughlin}). In both graphs, the system size is $100\times 100$ and the hopping terms are $r=t=1$.}
\end{figure}

We plot  $\zeta^{(g)}_H$ as a function of mass $m$ for various values of the hopping parameter $t$ in Fig.~\ref{fig:CI}(a). Since in this formalism we have a conserved momentum charge, we can compare this result to the Streda formula and Laughlin gauge argument calculations mentioned in Section II. In Fig. \ref{fig:CI_comp}(a) we show the momentum density in a cylinder geometry as a function of the geometric deformation for use in the generalized Laughlin experiment (c.f. Fig.~\ref{fig:spec_flow}). We see that the ``momentum charge'' with opposite signs is accumulated at the two edges of the cylinder as the viscoelastic magnetic flux $\phi_G$ is gradually dialed up. One can then read off the Hall viscosity using Eq.~(\ref{eq:Laughlin}).
We also compute $\zeta^{(g)}_H$ by the Streda formula in Eq.~(\ref{eq:Streda}). 
The outcomes of these calculations are plotted in Fig.~\ref{fig:CI_comp}(b), which shows that the Streda formula, as well as the Laughlin experiment, yield the same results as the Kubo formula. This remarkable agreement means that the concept of momentum charge in the gauge coupling formalism is still valid even in lattice models.

Next, we derive the viscoelastic response using the electron-phonon coupling formalism. 
According to Eq.~(\ref{eq:eph_hop}), the hopping matrix elements are modified such that
\begin{align*}
t_1\sigma_1 & \to t(1- u_{11})\sigma_1 + t u_{21}\sigma_2, \nonumber \\
t_2\sigma_2 & \to t(1- u_{22})\sigma_2 + t u_{12}\sigma_1, \nonumber \\
r_x &\to r(1-u_{11}), \nonumber \\
r_y &\to r(1-u_{22}),
\end{align*}
where we have used the approximation $\partial t(r)/\partial r\approx -t/a$~\footnote{For sufficiently localized orbitals in a tight-binding model, the hopping matrix element depends exponentially on distance $r$ between two orbitals; i.e.~$t(r)\propto e^{-r/a}$ where $a$ is approximately the equilibrium lattice constant.}. Because of this simplification, and the form of the Dirac model coupled to strain, there will not be an extra pre-factor in the final expression of the Hall viscosity like what we saw for the Hofstadter model. 
From this we find the corresponding stress currents:
\begin{align*}
\hat{\cal T}^{(p)}_{11}=& \frac{1}{2} \sum_{\textbf{x}} {\Big[} c_{\textbf{x}+\textbf{a}_1}^\dagger (i t \sigma_1 - r \sigma_3)  c_\textbf{x}   {\Big]}, \\
\hat{\cal T}^{(p)}_{12} =& \frac{it}{2} \sum_{\textbf{x}} {\Big[} c_{\textbf{x}+\textbf{a}_2}^\dagger  \sigma_1   c_\textbf{x} +c_{\textbf{x}+\textbf{a}_1}^\dagger \sigma_2  c_\textbf{x} + \text{h.c.}  {\Big]} ,
\end{align*}
which can be written in reciprocal space as
\begin{align*}
{\cal T}^{(p)}_{11} (\textbf{k})&= \sigma_1 t\sin k_1a - \sigma_3 r\cos k_1a, \\
{\cal T}^{(p)}_{12} (\textbf{k})&= t (\sigma_2 \sin k_1a + \sigma_1 \sin k_2a).
\end{align*}
Plugging these expressions into the commutator of Eq.~(\ref{eq:Kubo_visc}) leads to (see Appendix A for details of the derivation):

\begin{widetext}
\begin{align}  \label{eq:CI_phonon}
\zeta^{(p)}_H= \frac{1}{L^2} \sum_{\textbf{k}} {\cal B}^{(p)} (\textbf{k}) = \frac{1}{4Na^2}  \sum_{\textbf{k}} \frac{t^2 (\sin^2 k_2a\ (m - 2r \cos k_1a)+\sin^2 k_1a\ (m - 2r \cos k_2a))}{\left[ (m - r\cos k_1a - r \cos k_2a)^2 + t^2 (\sin^2 k_1a + \sin^2k_2a)\right]^{3/2}}. \\
\nonumber
\end{align}
\end{widetext}

The results of the above expression are plotted in Fig.~\ref{fig:CI}(b) as a function of $m$ and $t$. As we see in this figure, there is no discontinuity around the critical point ($|m|=2r$) from the topological to the trivial phases. This is in contrast with what was previously found in Ref.~\onlinecite{Maissam2012}. Our calculation (details in Appendix A) shows that there is a factor of two for the ``$r\cos k_i a$'' terms inside the parentheses in the numerator which is absent in the original calculation of Ref.~\onlinecite{Maissam2012}. The difference in the results is attributed to this factor alone. If the factor is absent then the continuum limit of the Hall viscosity close to the phase transition ($m\to 2r$) is different from Eq.~(\ref{eq:CI_cont_final}) (as we will see) and the leading order would be linear in the mass gap, $m-2r.$ Inclusion of the factor of two, as our calculation indicates, yields results which match the regularized continuum model  calculations of Refs.~\onlinecite{Taylor2011,Taylor2013} that find that the lowest order of the mass gap entering the viscosity coefficient is quadratic.

To gain a better understanding of what we have calculated, we look at the Hamiltonian in the continuum limit. Let us start from Eq.~(\ref{eq:Chern_gauge}). The electron operator can be written as
\begin{align*}
c_{\textbf{x}}= \sum_i e^{i\textbf{K}_i\cdot \textbf{x}} \psi_i(\textbf{x})
\end{align*}
where the slowly varying fields $\psi_i(\textbf{x})$, $i=0,1,2,3$, are introduced around the four Dirac points at $(0,0)$, $(\pi,0)$, $(0,\pi)$, and $(\pi,\pi)$. The long-wavelength effective theory near each possible Dirac point can be written as \begin{align}
\hat{H}^{(\text{cont.})} = \sum_{i,\mu,\nu} \psi_i^\dagger [v_F \alpha_{i,\mu} \sigma^\nu e^\mu_{i,\nu} \partial_\mu+ M_i \sigma_3] \psi_i
\end{align}
where the Fermi velocity $v_F=ta$, the masses are $\{M_i\}=\{ m-2r,m,m,m+2r\}$, the sign coefficients are $\{\alpha_{i,1}\}=\{+,-,+,-\}$ and $\{\alpha_{i,2}\}=\{+,+,-,-\},$ and $e^\mu_{i,\nu}$ is the frame field which can be written in terms of the distortion $w$
\begin{align*}
e^\mu_{i,\nu}= \delta^\mu_\nu-\alpha_{i,\mu} w_{\mu \nu} 
\end{align*}\noindent to linear order in $w.$

This is identical to the Hamiltonian considered in Ref.~\onlinecite{Taylor2011} using Pauli-Villars regularization. Therefore, the total Hall viscosity can be expanded as contributions from the neighborhood of each Dirac point
\begin{align*}
\zeta^{(\text{cont.})}_H \approx \sum_i C_i I(M_i)
\end{align*}
where $\{ C_i\}=\{ \alpha_{i,1}\alpha_{i,2}\}= \{+,-,-,+\}$ and
\begin{align} 
I(M) &= \frac{v_F^2}{16\pi^2} \int \frac{k^2 M}{(v_F^2 k^2+M^2)^{3/2}}d^2\textbf{k} \nonumber \\
&= \frac{M}{8\pi} \frac{2M^2+v_F^2 k^2}{v_F^2 \sqrt{v_F^2 k^2+M^2}}{\Big|}_0^\Lambda \nonumber \\
\label{eq:CI_cont_cutoff}
&= \frac{M \Lambda}{8\pi v_F} - \frac{M |M|}{4\pi v_F^2} 
\end{align}
where a phenomenological cut-off $\Lambda$ near each Dirac point has been introduced. It is worth mentioning that we have not set the sign coefficients $C_i$ arbitrarily as they appear in our calculations as a result of the long-wavelength expansion around each Dirac point. Hence,
\begin{align} \label{eq:CI_cont_final}
\zeta^{(\text{cont.})}_H \approx&\begin{cases}
    \frac{\hbar}{2\pi v_F^2} (-m|m|+ 4mr ) & |m|<2r \\
    \frac{2\hbar}{\pi v_F^2} r^2 & |m|\geq2r 
  \end{cases}
\end{align}
in which the (arbitrary) $\Lambda$-dependent term has vanished.  

For comparison, if we take the continuum limit of the expression derived from electron-phonon coupling, Eq.~(\ref{eq:CI_phonon}), this leads to the same results. For example.  near the critical point, $m=2r+\varepsilon$, the difference in $\zeta_H$ between trivial and topological phases is 
\begin{align}
\Delta\zeta_H^{(\text{cont.})}= \frac{\hbar}{2\pi v_F^2}\varepsilon^2 
\end{align}
which is the same as the regularized field theory result~\cite{Taylor2011}.

\begin{figure}
\includegraphics[scale=0.5]{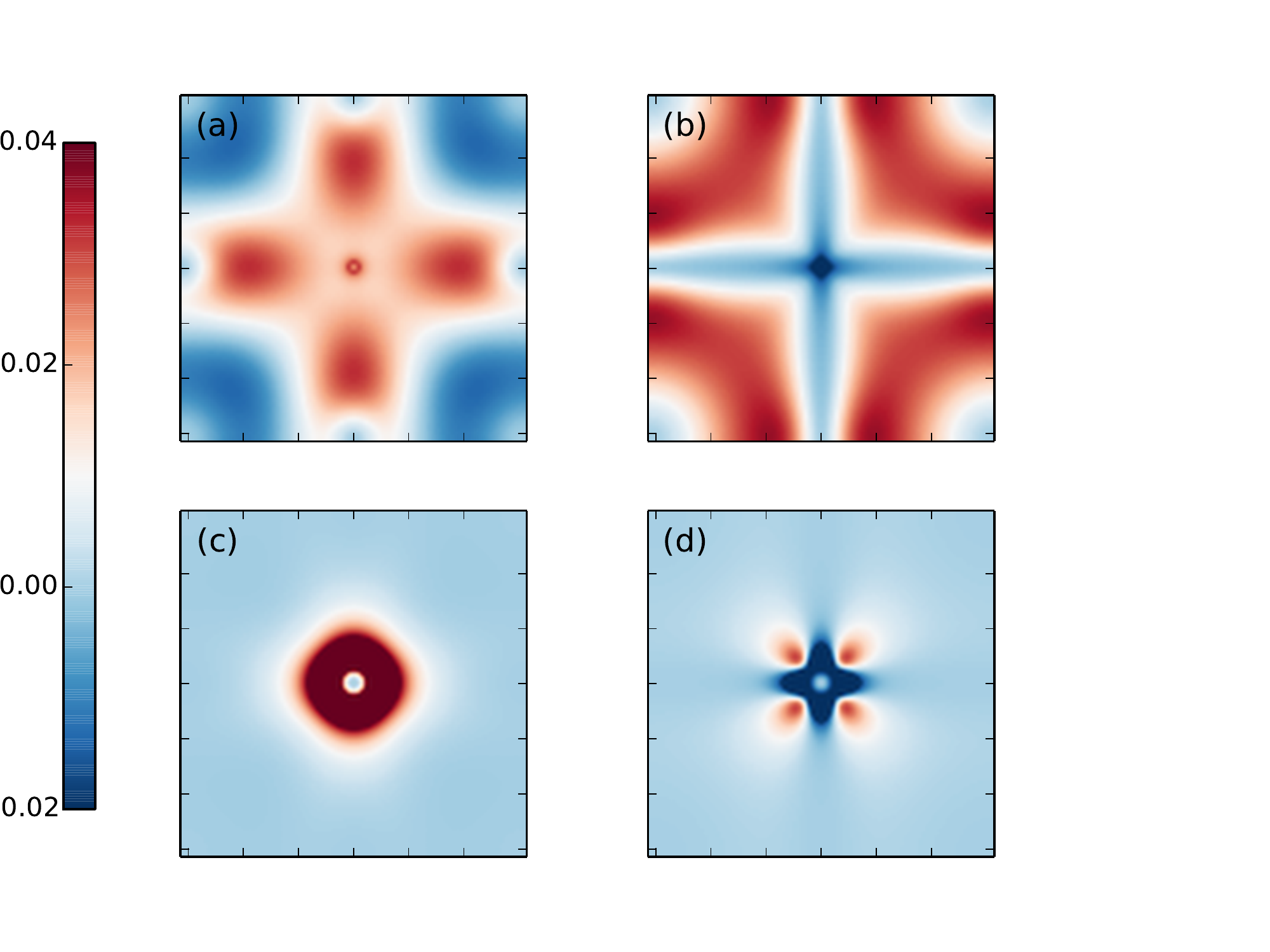}
\caption{\label{fig:Berry}  The adiabatic curvature over the Brillouin zone in the vicinity of the critical point  ($m-2r=-0.1$). (a) and (c) are computed by Eq.~(\ref{eq:CI_gauge}) and (b) and (d) are computed by Eq.~(\ref{eq:CI_phonon}). In (a) and (b), $t=2$ and in (c) and (d), $t=0.25$. The energy scale is set by $r=1$. The grid in reciprocal space is $200\times 200$.}
\end{figure}

A few remarks about Fig.~\ref{fig:CI} are in order. We have also included the results of momentum polarization method~\cite{P_pol_Hvisc,DMRG_Hvisc} in panel (c) for reference. The Hall viscosity is an odd function of the mass parameter $m$; this is consistent with the fact that the chirality is flipped as the mass changes sign for our model. However, the Hall viscosity does not vanish in the trivial phase as it does in the regularized field theory calculation. Interestingly, we do find that as the hopping coefficient $t$ becomes smaller (i.e., $v_F\to 0$) $\zeta_H$ goes to zero much more rapidly away from the critical point and into the trivial phase. 
Using these results we can interpret the discrepancy about the residual viscosity in the trivial phase from several viewpoints. 
First, from a symmetry point of view, this result is not in contradiction with the requirements of generically having a non-zero Hall viscosity since  time reversal symmetry is broken everywhere in the phase diagram. While the Chern number also requires time-reversal breaking, it does vanish in the trivial phase which is natural since the Chern number must be quantized, and is thus much more constrained. Moreover, crudely speaking, as we decrease $t$ in the regime $t<r$, the effective bulk gap becomes smaller and the corresponding correlation length becomes larger compared with the lattice constant. This essentially makes the lattice effects less important; consequently, we would expect $\zeta_H$ to  more closely match  the continuum limit results where the Hall viscosity  vanishes in the trivial phase.

\begin{figure}
\includegraphics[scale=0.8]{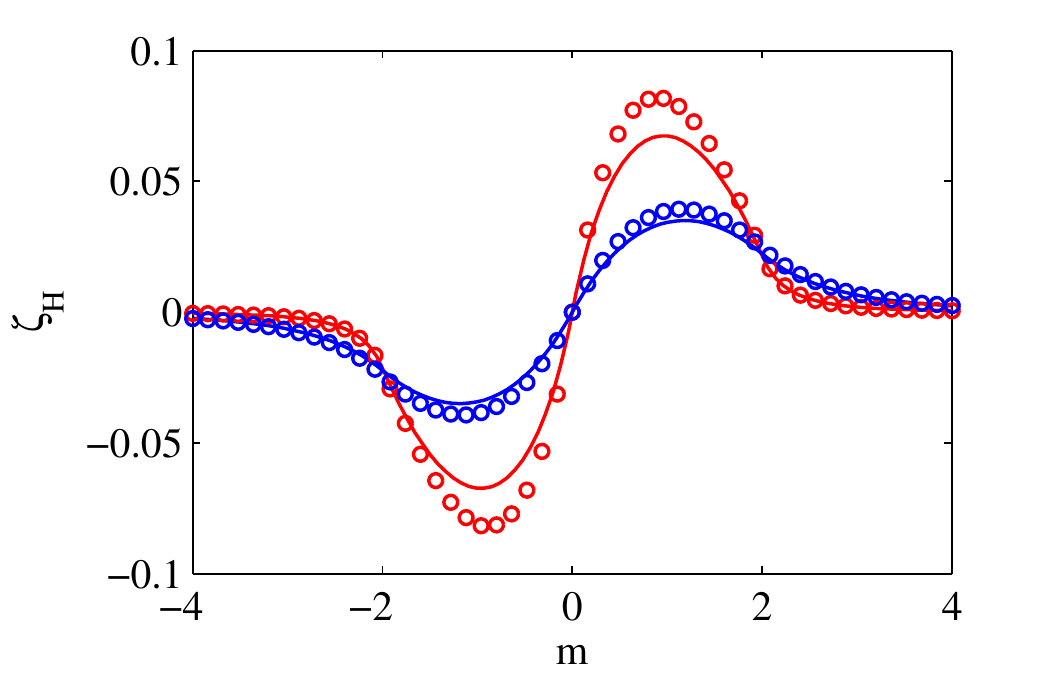}
\caption{\label{fig:pol_comp}  Comparison of the Hall viscosity using the gauge formalism in Eq.~(\ref{eq:CI_gauge}) (solid) and the momentum polarization method~\cite{P_pol_Hvisc,DMRG_Hvisc}(circles). $t=2.0$ (blue) and $t=0.75$ (red). }
\end{figure}

We can also understand the residual viscosity from another point of view by considering the viscoelastic adiabatic functions ${\cal B}^{(g/p)}(\textbf{k})$ defined in Eqs. (\ref{eq:CI_gauge}) and (\ref{eq:CI_phonon}). For moderate regimes of $t$ (when $t\sim r$)  the quantities ${\cal B}^{(g/p)}(\textbf{k})$  spread over a wider region of the Brillouin zone (Fig.~\ref{fig:Berry}(a) and \ref{fig:Berry}(b)) while for smaller values of $t$, they become localized around the origin (Fig.~\ref{fig:Berry}(c) and \ref{fig:Berry}(d)). Hence it is the latter case where we expect the continuum expansion to become more accurate, i.e., where this function is only sampling the contribution in the neighborhood of a Dirac point.

Finally, from numerical point of view, the Hall viscosity is given in terms of the complex phase structure of the single-particle wave functions. If one of the three Pauli matrices was absent from the Hamiltonian, the wave function could be made completely real. In fact, deep in the trivial phases  the $\sigma_3$ term is dominant over the entire Brillouin zone; hence all the wave functions are almost real in this regime and the Hall viscosity is negligible. In the vicinity of the transition point ($|m|=2r$), but on the trivial side, the $\sigma_1$ and $\sigma_2$ terms can still be comparable with the $\sigma_3$ term within some regions of the Brillouin zone and this contributes to a non-zero Hall viscosity. As we tune down the hopping amplitude $t$, i.e., the amplitude of the   $\sigma_1$ and $\sigma_2$ terms, the region of the Brillouin zone over which these terms are comparable to the $\sigma_3$ term shrinks, and $\zeta_H$ asymptotes to zero faster. Hence because of the more complicated nature of the viscosity it is natural for it to be non-vanishing, even in the trivial phase. However, deep in the trivial phases, i.e., where we could consider the system in a trivial atomic limit, the viscosity indeed vanishes. 

It is interesting to note that besides the fact that the results of momentum polarization (Fig.~\ref{fig:CI}(c)) satisfy all the aforementioned properties, it yields similar curves as the gauge formalism, a comparison is shown in Fig.~\ref{fig:pol_comp}. Remarkably, the difference between two methods are minimized as we approach the critical regions where the lattice effects are small.

\section{Example 3: surface states of the 3D topological insulator}

For our final model we study the standard Wilson-Dirac Hamiltonian on a cubic lattice as a simple model of the 3D TI~\cite{Wilson_3DTI,Qi_3DTI}
\begin{align} \label{eq:3DTI}
\hat{H}=& \frac{1}{2} \sum_{\substack{\textbf{x}\\ s=1,2,3}} {\Big[} c_{\textbf{x}+\textbf{a}_s}^\dagger (i t_s \alpha_s - r \beta) c_\textbf{x} +\text{h.c.} {\Big]} \nonumber \\ &+ (m+3r) \sum_\textbf{x} c_\textbf{x}^\dagger \beta c_\textbf{x}
\end{align}
where the Dirac matrices are given by
\begin{align*}
\alpha_s&= \tau_1\otimes \sigma_s=\left(\begin{array}{cc}
0 & \sigma_s \\ \sigma_s & 0
\end{array} \right), \nonumber \\
\beta&= \tau_3\otimes 1=\left(\begin{array}{cc}
\mathbbm{1} & 0 \\ 0 & -\mathbbm{1}
\end{array} \right), \nonumber \\ 
\gamma_5&= \tau_1\otimes \sigma_1=\left(\begin{array}{cc}
0 & \mathbbm{1} \\ \mathbbm{1} & 0
\end{array} \right) .
\end{align*}
For our calculations we will take the Wilson parameter to always be fixed at $r=1$, and the system is taken to have cubic symmetry with $t_1=t_2=t_3=t.$ 
In this convention the $\sigma$ and $\tau$ matrices act on the spin and orbital degrees of freedom respectively.

Transforming to reciprocal space, the Bloch Hamiltonian reads
\begin{align*}
h(\textbf{k})= \sum_{s=1,2,3}{\Big[} t_s \alpha_s \sin k_s - r\beta \cos k_s {\Big]}+ (m+3r)\beta .
\end{align*}
There are two important symmetries present in this model:(i) time-reversal symmetry (TRS) with the operator ${\cal T} = i \sigma_2 {\cal K}$ such that $\sigma_2 h(\textbf{k})\sigma_2= h^\ast(-\textbf{k})$; (ii) inversion symmetry (IS), represented by ${\cal I}= \tau_3{\cal P}$, such that $\tau_3 h(\textbf{k})\tau_3=h(-\textbf{k})$. This model can exhibit a non-trivial 3D TI phase protected by time-reversal symmetry.
In fact, as the mass parameter $m$ is varied, the Hamiltonian shows the following phases:
\begin{enumerate}[(a)]
\item
$0 < m$ and $m <-6r$: trivial phase equivalent to the atomic limit.
\item
$-2r<m<0$ and $-6r<m<-4r$: strong TI with a single Dirac cone on each boundary surface.
\item
$-4r<m<-2r$ : weak TI with an even number of Dirac cones on each boundary surface.
\end{enumerate}

Here, we are only interested in the strong TIs in case (b) which have surface states  described by a single Dirac Hamiltonian. For these phases it has been predicted that opening a gap in the surface states with a magnetic layer will induce a half-quantized quantum Hall effect\cite{Qi_3DTI,Moore-Vanderbilt,Nomura_ME} and an accompanying surface Hall viscosity\cite{Taylor2011,Onkar}. While the former prediction has been carefully studied in the context of the topological magneto-electric effect (also known as axion electrodynamics), the latter prediction has not been confirmed in a physical lattice model. Hence, our goal in this section is to explore the surface Hall viscosity response, and to also search for a route for experimental measurement by calculating the viscosity-modified surface phonon dispersion relations, as well as detecting non-local phononic responses (see Ref. \onlinecite{Maissam2012} for similar phonon calculations for 2D systems).

To calculate the viscosity we consider a slab geometry with a finite number of layers along the open-boundary $z$-direction, we keep the $x$ and $y$ directions translation invariant and periodic. We will induce a surface Hall viscosity by breaking TRS on the boundaries via TR breaking Zeeman terms on the upper/lower surfaces
\begin{align*}
\hat{H}_{FM}= \sum_{\substack{\textbf{x}=(x,y)\\ z=0,L_z}}  c_{\textbf{x},z}^\dagger\ \Omega_z \sigma_3 \ c_{\textbf{x},z} .
\end{align*}
This term can be interpreted microscopically as layers of a ferromagnet deposited on the upper/lower surfaces that produce a magnetic field strength $\Omega_z$. We take $\Omega_0=-\Omega_{L_z}=\Omega$ to have opposite signs, although this choice is not crucial, i.e., choosing the same sign yields identical results in our geometry though it might lead to complications in the presence of side surfaces. The ferromagnetic term  opens a Zeeman gap in the spectrum of topological surface states
\begin{align}
H_{u (l)} = \sigma_1 k_1 \mp \sigma_2 k_2 \pm \Omega \sigma_3
\end{align}
where $u(l)$ subscripts refer to upper(lower) surface states and are correlated with the signs. This Hamiltonian is valid in momentum space near the $\Gamma$-point up to a momentum cut-off $\Lambda$, which depends on the bulk mass parameter $m,$ and the hopping parameter $r.$ 
  
  The gapped Dirac surface states on each surface have a Hall conductance of $\sigma_H=e^2/2h$ which  has been directly computed using several methods in Refs.~\onlinecite{Qi_3DTI,Moore-Vanderbilt,Nomura_ME,Franz_Witten_effect}.
To calculate the viscosity, we can generalize one of the methods of Ref. \onlinecite{Moore-Vanderbilt}, the ``layer-resolved Chern number'', to compute the Hall viscosity layer by layer instead. The Hall conductance for a slab geometry can be written as $\sigma_H= Ce^2/h,$ where $C$ is the integer Chern number~\cite{TKNN1982} given by
\begin{align*}
C = \frac{ 2\pi}{i L^2} \sum_{\textbf{k}}  \text{Tr}{\big[} {\cal P}_{\textbf{k}} \epsilon_{ij} (\partial_i {\cal P}_{\textbf{k}}) (\partial_j {\cal P}_{\textbf{k}}) {\big]}
\end{align*}
in which  ${\cal P}_{\textbf{k}}=\sum_\nu |u_{\nu\textbf{k}}\rangle \langle u_{\nu\textbf{k}}|$ is the projection operator onto the occupied states of the Hamiltonian in the slab geometry, and $\partial_i = \frac{\partial}{\partial k_i}$. To find how different $z$ layers contribute to $C$ we can define a projection operator $\tilde{\cal P}_z= |z\rangle \langle z|$ onto the $z$-th layer and compute
\begin{align*}
C = \frac{ 2\pi}{i L^2} \sum_{\textbf{k}}  \text{Tr}{\big[} {\cal P}_{\textbf{k}} \epsilon_{ij} (\partial_i {\cal P}_{\textbf{k}}) \tilde{\cal P}_z (\partial_j {\cal P}_{\textbf{k}}) {\big]}.
\end{align*}

\begin{figure}
\includegraphics[scale=0.8]{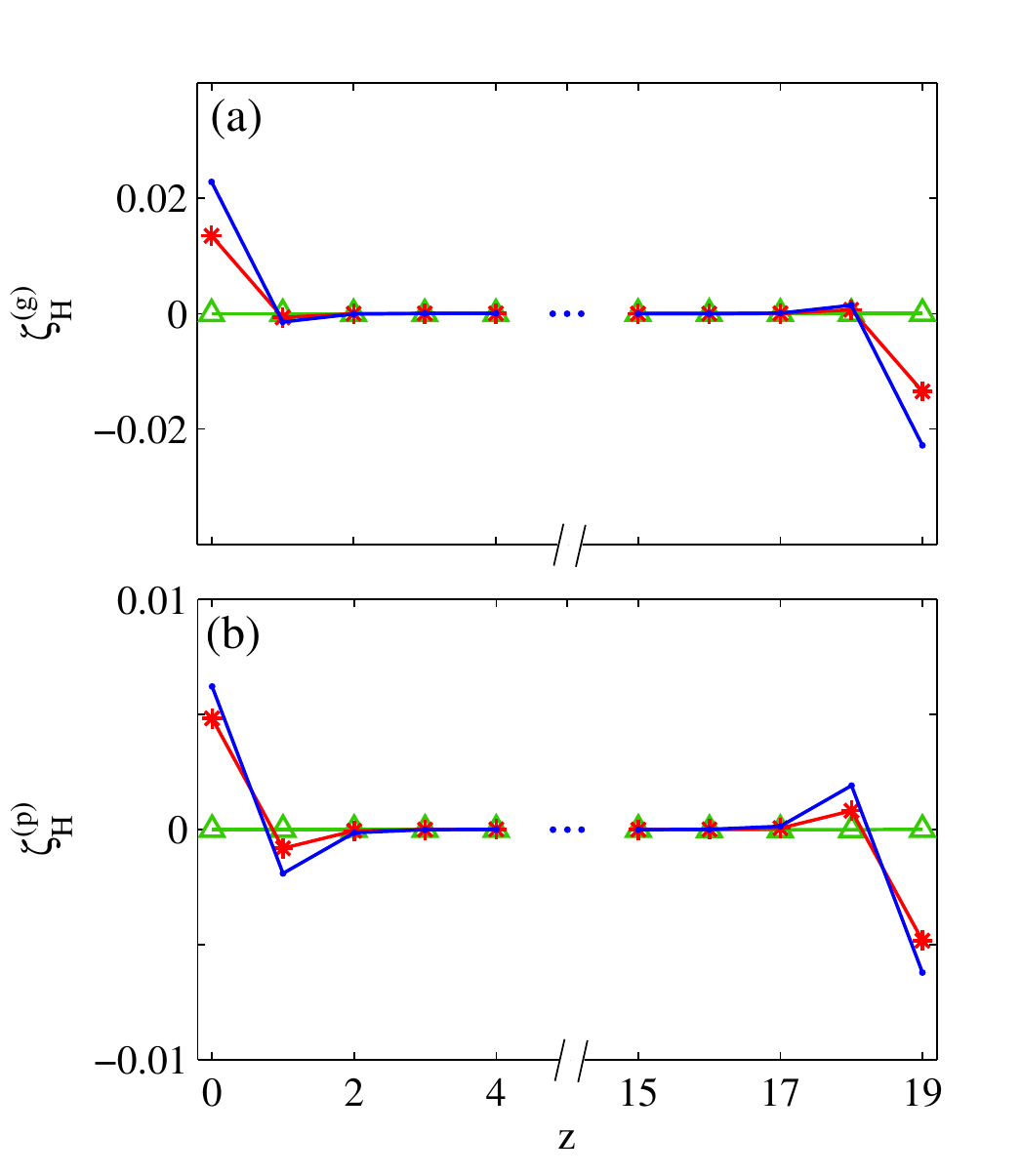}
\caption{\label{fig:3D_profile}  Layer-resolved Hall viscosity using (a) the gauge coupling and (b) the electron-phonon coupling methods. Different colors represent different values of the surface gap: $\Omega= 0.0$(green, triangles), $0.2$(red, stars), and $0.4$(blue, circles). Other parameters are: $m=-1$,  $r=t=1$, and $(L_x,L_y,L_z)=(100,100,20)$}
\end{figure}

\begin{figure}
\includegraphics[scale=0.8]{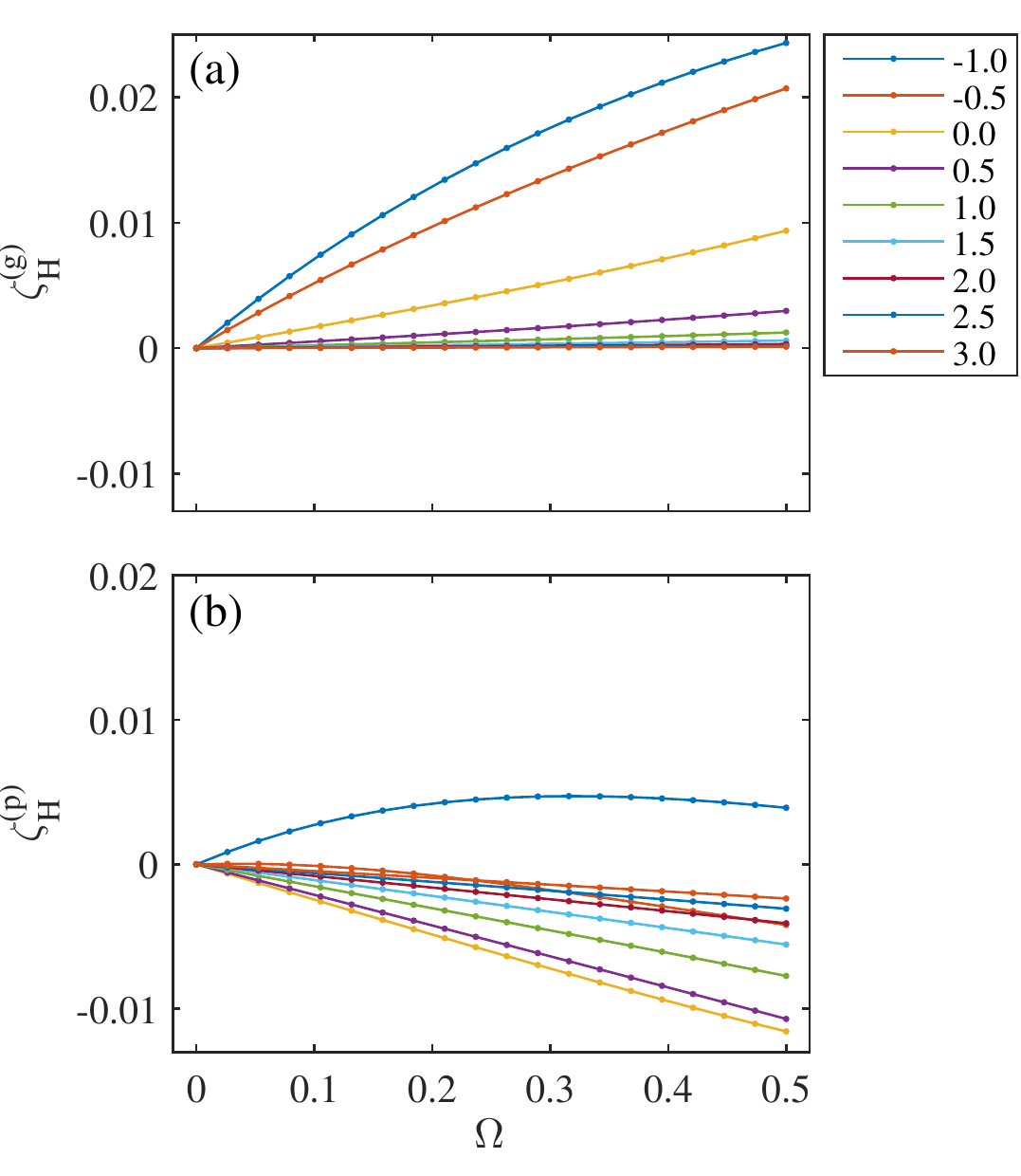}
\caption{\label{fig:HV_vs_b}  The Hall viscosity of the surface states of 3D TI. (a) The gauge coupling method Eq.~(\ref{eq:3D_gauge}). (b) The electron-phonon coupling method Eq.~(\ref{eq:3D_phonon}). The solid lines are fits using Eq.~(\ref{eq:fit_exp}). The legend shows the bulk mass $m$ values. The hopping parameters are $t=r=1$. The system size is $(L_x,L_y,L_z)=(100,100,20)$.}
\end{figure}

\begin{figure}
\includegraphics[scale=0.8]{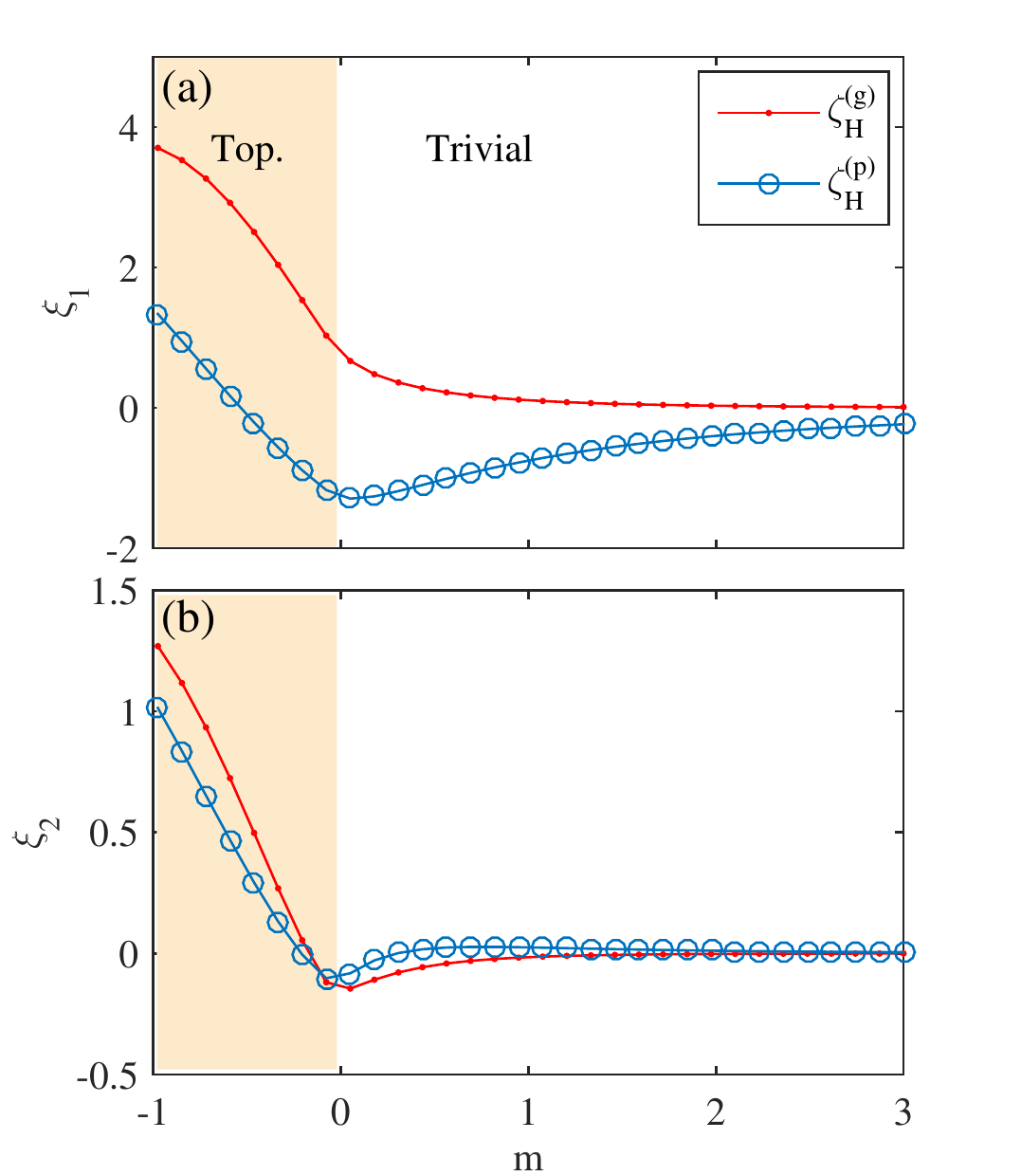}
\caption{\label{fig:3D_coefs}  The coefficients of the fitting expression, Eq.~(\ref{eq:fit_exp}), for the Hall viscosity in Fig.~\ref{fig:HV_vs_b}. Notice the coefficient of the quadratic term ($\xi_2$) is much more method-independent than the linear term ($\xi_1$). Parameter details at Fig.~\ref{fig:HV_vs_b}.}
\end{figure}

The same idea can be applied to the Hall viscosity. Starting from Eq.~(\ref{eq:proj_visc_gauge}), the gauge coupling formula, one can calculate the contribution of layer $z$ to the Hall viscosity as 
\begin{align} \label{eq:3D_gauge}
\zeta^{(g)}_H(z)= \frac{1}{2i L^2} \sum_{\textbf{k}} \ &(\sin^2k_1+\sin^2 k_2) \nonumber \\
&\times \text{Tr}{\big[} {\cal P}_{\textbf{k}} \epsilon_{ij} (\partial_i {\cal P}_{\textbf{k}}) \tilde{{\cal P}}_z (\partial_j {\cal P}_{\textbf{k}}) {\big]}.
\end{align}
Similarly, one can write down a layer-resolved expression for the Hall viscosity based on  the electron-phonon coupling formula in Eq.~(\ref{eq:proj_visc_ph}),
\begin{align} \label{eq:3D_phonon}
\zeta^{(p)}_H(z)= \frac{1}{iL^2}  \sum_{\textbf{k}} \ \text{Tr}{\big[} {\cal P}_{\textbf{k}} \epsilon_{ij} (\partial_{u_{1i}} {\cal P}_{\textbf{k}}) \tilde{{\cal P}}_z (\partial_{u_{2j}} {\cal P}_{\textbf{k}}) {\big]}.
\end{align}
Figure \ref{fig:3D_profile} shows the layer-resolved Hall viscosity along the open boundary $z$-direction using both methods. Next, the total Hall viscosity is calculated by summing over the contributions from half of the lattice.

The results of both methods for various values of $\Omega$ are shown in Fig.~\ref{fig:HV_vs_b}. As we see, regardless of the bulk mass parameter $m$, the Hall viscosity is zero when the surface states are gapless $\Omega=0$. 
Inspired by the continuum limit expression for the single Dirac cone in Eq~(\ref{eq:CI_cont_cutoff}), we can fit the data with the following ansatz 
\begin{align} \label{eq:fit_exp}
\zeta_H= \frac{\xi_1(m)}{8\pi v_F} \Omega - \frac{\xi_2(m)}{4\pi v_F^2} \Omega^2 \text{sign}(\Omega)
\end{align}
where the coefficients $(\xi_1,\xi_2)$ depend on the bulk TI gap.
The sign change in curvature (sign of the quadratic term) as we pass from $\Omega>0$ to $\Omega <0$ is confirmed by our calculations.
As shown in Fig.~\ref{fig:3D_coefs}, both coefficients $(\xi_1,\xi_2)$ decrease as the bulk mass is swept towards the trivial phase. From Eq.~(\ref{eq:CI_cont_cutoff}), and the fact that the cutoff is proportional to the bulk mass, the linear term should vanish as the bulk gap closes ($m\to 0$); while as we see in Fig.~\ref{fig:3D_coefs}(a), $\xi_1$ stays non-zero beyond $m=0$ and within the bulk trivial phase. This effect is similar to the one observed in 2D case where the Hall viscosity gradually vanishes in the trivial phase. We check that decreasing $t$ indeed makes $\xi_1$ transition to zero faster.
In addition, the regularized 2D continuum expression in Eq.~(\ref{eq:CI_cont_cutoff}) implies that the quadratic term should not depend on the cutoff; therefore, we might expect that $\xi_2=1$ and should remain constant in the topological phase. However, our lattice calculations in Fig.~\ref{fig:3D_coefs}(b) show that $\xi_2$ is close to $1$ only at the deepest point in the topological phase, and decays to zero as the bulk critical point is approached. We find that the coefficients of the quadratic term from both methods show similar magnitudes and trends, while the behavior of the nominally cut-off dependent coefficient $\xi_1(m)$ seems quite method dependent as we might expect from the continuum calculations.

Now that we have shown that a Hall viscosity can appear at the surface of a 3D TI, we will use the electron-phonon coupling formalism to discuss that the Hall viscosity can add new phononic properties to the TI, which could be experimentally measurable. 
To this end, let us start with the effective theory for the phonon field.
The phonon dynamics in the harmonic limit are governed by the following Lagrangian density
\begin{align*}
{\cal L}_{\text{ph}}= \frac{1}{2} {\Big(} \rho \dot{u_i}^2 - c_{ijkl} u_{ij}u_{kl} {\Big)},
\end{align*}
where $\rho$ is the mass density and, for simplicity, we consider an intrinsically isotropic system where the elasticity tensor takes the form $c_{ijkl}=\lambda \delta_{ij} \delta_{kl} + \mu (\delta_{ik}\delta_{jl} + \delta_{il}\delta_{jk})$. The parameters $\lambda$ and $\mu$ are called Lam\'{e} constants~\cite{Landau1986}.

According to the linear response theory, Eq.~(\ref{eq:Kubo_visc}), the extra term due to the Hall viscosity is
\begin{align}
{\cal L}^{2D}_{\text{visc}}=  \zeta_H \epsilon_{ij} \ \dot{u}_{ni} u_{nj}
\end{align}
for a 2D Chern Insulator and,
\begin{align}
{\cal L}^{3D}_{\text{visc}}=  \zeta_H \epsilon_{ijk} \ \dot{u}_{ni} \partial_j u_{nk}
\end{align}
for a 3D TI, which is an analog of the magneto-electric term $E\cdot B$.

We shall discuss the 2D case (which was originally explained in Ref.~\onlinecite{Maissam2012}) as a warm up example. The equation of motion is found to be
\begin{align}\label{eq:EOM_2D}
\ddot{u}_i=& v_s^2 \del^2 u_i  + (v_\ell^2 - v_s^2)\partial_i \del\cdot\textbf{u} + \epsilon_{ij} \zeta_H \del^2 \dot{u}_j /\rho 
\end{align}
where the velocities of the longitudinal (LA) and transverse (TA) acoustic phonons are defined via $v_\ell=\sqrt{(\lambda+2\mu)/\rho}$ and $v_s=\sqrt{\mu/\rho}$, respectively. The Hall viscosity term couples the LA and TA modes and modifies the phonon dispersion relation as 
\begin{align} \label{eq:disp_2D}
\omega&\approx v_\alpha k + c_\alpha  \left(\frac{\omega}{\omega_{H}}\right)^2 (\frac{v_\ell^2-v_s^2}{v_\alpha})\ k,
\end{align}
for different modes  $\alpha=\ell$ or $s$, where the characteristic frequency determined by the Hall viscosity is $\omega_{H}=\rho (v_\ell^2-v_s^2)/ \zeta_H,$ and $c_{\ell(s)}=\pm 1$. The other consequence of the Hall viscosity term is that if we drive mode $\alpha$ with amplitude $A_\alpha$, the other mode $\bar{\alpha}$ will acquire a non-zero relative amplitude of
\begin{align*}
\frac{A_{\bar{\alpha}}}{A_\alpha}
= i \frac{\omega}{\omega_{H}} .
\end{align*}
Notice that the Hall viscosity generated phonon wave  has  a $\pi/2$ phase shift which may make  measurements easier.

Now let us consider a 3D TI. For our 3D system let us assume a slab geometry with finite length along the $z$-direction.
The equation of motion reads
\begin{align}\label{eq:EOM_3D}
\ddot{u}_i=& v_s^2 \del^2 u_i  + (v_\ell^2 - v_s^2)\partial_i \del\cdot\textbf{u} + \epsilon_{ijk} (\partial_j \zeta_H) \del^2 \dot{u}_k /\rho. 
\end{align}
Note that unlike Eq.~(\ref{eq:EOM_2D}), the Hall viscosity term here, $\nabla \zeta_H(z) = \zeta_H\ \delta(z) \hat{z}$, is a boundary term that modifies the surface phonon dispersion, as well as the boundary conditions.
The correction to the dispersion relation (now only at the surface) has the same form as Eq.~(\ref{eq:disp_2D}) up to replacing the 2D density $\rho$ with the 3D density multiplied by the penetration depth of the surface states (roughly few layers in the $z$-direction, e.g. a few nm in Bi$_2$Se$_3$), $\rho \to \rho \xi$ where $\xi\sim \hbar v_F/E^{(bulk)}_g$ is the surface state penetration depth. In addition, we discuss two unique signatures of the Hall viscosity in a 3D geometry: the analog of the Faraday rotation for normally incident TA phonons, and the emergence of a TA phonon wave traveling away from the surface due to the excitation of  surface LA phonons.

First let us consider the phonon Faraday rotation. The general form of TA phonons propagating along the open boundary $z$-direction is given by the displacement field $\textbf{u}(z)= u_1(z) \hat{x} + u_2(z) \hat{y}$. Suppose TA phonons in the $x$-direction are excited, $u_1(z)= A_1  e^{i (\omega/v_s) z-i\omega t}$.  The boundary condition for the other TA mode is
\begin{align*}
\partial_z u_2{\Big|}_{z=0+} = - i \frac{\zeta_H}{\rho v_s^4} \omega^3 u_1(z=0).
\end{align*}
Note that in absence of the Hall viscosity, this equation reduces to the boundary condition for the free surface: $\partial_z u_2=0$ and $u_2$ remains zero. However, for $\zeta_H\neq 0$, this mode is generated, $u_2= A_2  e^{i (\omega/v_s) z-i\omega t}$, with a relative amplitude of
\begin{align*}
\frac{A_2}{ A_1}= \frac{\zeta_H}{\rho v_s^3} \omega^2.
\end{align*}
Hence the polarization of the phonon is rotated by a small angle $\theta$ given by $\tan \theta =A_2/A_1$ in the $xy$-plane. The rotation of oscillation axis is an analog of the Faraday rotation for the TA phonons. 

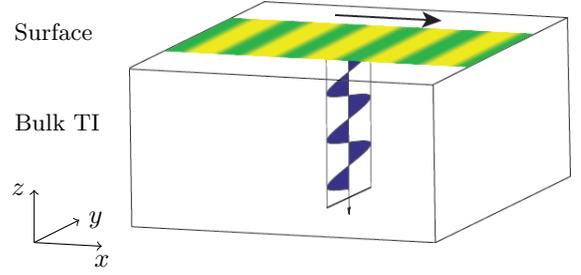
\begin{figure}
\begin{tikzpicture}
\node[anchor=center,inner sep=0] at (0.5,-0.2) {\includegraphics[scale=0.35]{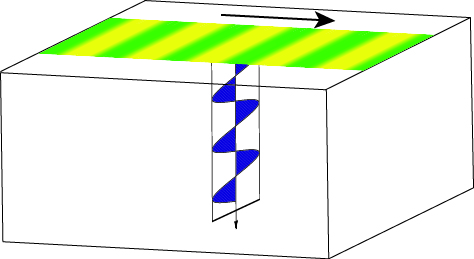}};
\draw (-2.8,1) node[left] {\small Surface};
\draw (-2.7,-0.2) node[left] {\small Bulk TI};
\draw[->] (-3.7,-1.8)--(-2.8,-1.85) node[below]{ $x$};
\draw[->] (-3.7,-1.8)--(-3.7,-1.1) node[left]{ $z$};
\draw[->] (-3.7,-1.8)--(-3.1,-1.5) node[right]{ $y$};
\end{tikzpicture}
\caption{\label{fig:SAW} Traveling TA phonon wave into the bulk TI as a result of driving the surface LA phonons. The green and yellow stripes are to represent compression and decompression regions respectively. Arrows show the propagation directions. }
\end{figure}

Next, let us now illustrate how boundary effects can lead to a non-trivial phononic effect (which is unique to phonons and does not exist for the photonic response) for a 3D TI. Consider the following general form for the phonon field $\textbf{u}(x,z)=u_1(x,z) \hat{x} + u_2(x,z) \hat{y}$. Suppose the surface phonons are driven in the LA mode as $u_1(x,z=0)=A_1 e^{i (\omega/v_\ell) x-i\omega t}$. The boundary condition for the TA mode can be derived as
\begin{align*}
& \partial_z u_2{\Big|}_{z=0+} = -i \frac{\zeta_H}{\rho v_s^2 v_\ell^2} \omega^3  u_1(z=0).
\end{align*}
This equation implies that there will be a phonon wave traveling away from the surface, given by the ansatz $u_2 (x,z)=A_2   e^{i (\omega/v_\ell) x+ik_3 z-i\omega t}$, where $k_3=\omega (v_\ell^2-v_s^2)^{1/2}/v_s v_\ell$ can be found from the wave equation in the bulk. The amplitude of such a mode can be computed from the boundary condition
\begin{align*}
\frac{A_2}{ A_1}= \frac{\zeta_H}{\rho (v_\ell^2 - v_s^2)^{1/2}v_s v_\ell} \omega^2.
\end{align*}
This phenomenon could be considered as a direct signature of the Hall viscosity in a 3D TI: the TA phonons can be detected on the bottom surface as a result of exciting the LA phonons on the top surface (illustrated in Fig.~\ref{fig:SAW}).

To be a bit more quantitative we estimate the correction due to the Hall viscosity as follows: for 2D systems, a typical solid has a mass density of $\rho\sim 10^{-7}\ g/cm^2$,  phonon velocities of $v_\ell \approx 2v_s \sim 10^5\ cm/s$, and  $\zeta_H \sim 10^{-1} \hbar/a^2$ as we have found in the previous section. Consequently, the characteristic frequency is approximately $\omega_{H} \sim 10^{15} s^{-1}$. The important assumption in our calculation is to treat the phonon in the adiabatic limit; i.e.,~$\omega\ll \omega_D, E_g$, where a generic value for the Debye frequency is $\omega_D\sim 10^{14} s^{-1}$, and the electronic bulk gap is $E_g \sim 10^{14} s^{-1}$. Hence, one can reach rather considerable ratios up to $\omega/\omega_H \sim 10^{-2}$ at frequencies around $1\ THz$. For 3D systems, $\rho\sim 10\ g/cm^3$ and  $\zeta_H \sim 10^{-2} \hbar/a^2$ as we have calculated in this section; therefore, the ratio $A_2/A_1$ can be made as large as $10^{-2}$.  Similar orders of magnitude, $10^{-2}$, can be accessed for the corrections to the surface phonon dispersion of 3D TIs. 

There are already various experimental techniques that might be used to investigate the unusual phononic properties of TIs due to the Hall viscosity. 
In terms of materials to be studied, $\text{Bi}_2\text{Se}_3$ as an archetypal 3D strong TI could be a promising choice, especially since recent experiments have shown that the electron-phonon coupling at its surface is strong~\cite{BiSe_He_spec}.
To study 2D effects (wave mixing), the surface acoustic (Rayleigh) waves method can be used to excite and measure different phononic modes. This method has been applied successfully to measure the electron-phonon coupling of a 2D electron gas (GaAs heterostructures) in integer QHE~\cite{SAW_wixforth}  and fractional QHE~\cite{SAW_willett} regimes. Recently, a similar setup was examined for the study of phonons coupled to Dirac electrons in graphene~\cite{SAW_graphene}.

To explore the 3D properties (analog of Faraday rotation, and emergence of a normal propagating mode), there are pump-probe measurements where LA/TA phononic modes are excited as a result of the relaxation of hot carriers photoexcited with ultrashort ($\sim100$ fs) laser pulses. For the slab geometry, normally incident phonons are directly measured using different techniques such as time-resolved x-ray diffraction~\cite{Pump_probe_xray,Pump_probe_xray_2}
 and transient reflectivity (TR) traces~\cite{BS_pump_probe_1,BS_pump_probe_2,BS_pump_probe_3}. The first method can be used to capture both LA and TA phonons while the second method is particularly useful for measuring the LA phonons. Recent experiments on phonon dynamics in $\text{Bi}_2\text{Se}_3$~\cite{BS_pump_probe_3} can in principle be readily adapted to study the Hall viscosity effects by adding ferromagnetic surface layers or applying an external magnetic field.
As a final remark, we note that crystal imperfections may also lead to mixing between LA and TA phonons, however, they can be distinguished from the Hall viscosity related effects because they do not depend on the sign of the time-reversal symmetry breaking terms such as the direction of magnetization in  ferromagnetic layers or external magnetic field.

In summary, from these calculations we can expect a non-zero surface Hall viscosity, but its exact value and dependence on the bulk mass $m$ is  potentially quite  complicated, and model dependent. We see some expected trends in the fitting coefficients that qualitatively match a continuum model for the surface states, and the proper symmetry constraints are met, but the results seem dependent on the choice of method, and on the high-energy regulator of the surface theory, i.e. the bulk mass $m.$  The presence of a non-zero Hall viscosity, regardless of the complicated dependence on details of the band structure, leads to various measurable effects in the phononic response of TIs: it modifies the phonon dispersion relation (bulk phonons in case of a 2D Chern insulator, and surface phonons in case of 3D) and it mixes orthogonal phonon modes which would be  otherwise independent.
There are other interesting questions one may ask about the emergent Hall viscosity on the boundaries of 3D TIs. Here, we computed the surface contributions due to the Zeeman term imposing the periodic boundary conditions on the boundaries of the top and bottom surfaces in a slab geometry; one can remove this boundary condition to see how gapless chiral domain wall states on side surfaces contribute to the Hall viscosity when there is a magnetization domain wall. Another possibility is that one can include orbital magnetic field terms, as a result of which Landau orbits will form in presence of a strong magnetic field. We speculate that the massive Dirac surface states would have an anomalous Hall viscosity due to the non-zero Dirac mass, in addition to the usual Hall viscosity coming from the Landau orbits.
As a slightly different direction, one can also study Hall viscosity purely due to 3D bulk when time-reversal is explicitly broken in the bulk. We expect that in this case viscoelastic response would be non-zero without any magnetic layers on boundary surfaces.



\section{Discussion}

We have presented two different schemes to implement lattice deformations in order to calculate the Hall viscosity in tight-binding models. We have validated these methods showing that, numerically and analytically, they both converge to the same results in the weak field limit of integer quantum Hall on the lattice (Hofstadter model).
Next, we have applied them to the 2D Chern insulator model and have demonstrated that, although different methods give rise to different values for the Hall viscosity, they satisfy certain common properties due to symmetry considerations; i.e. the Hall viscosity flips sign as the Chern number changes sign, vanishes at the particle-hole symmetric point $m=0$ where the Chern number changes sign, is continuous across the topological phase transitions, and asymptotes to zero in trivial phases. A comparison with the momentum polarization method shows that the gauge formalism matches well with the momentum polarization results near the phase transition points where the lattice effects are minimized.
We illustrated that confining the Berry curvature to the origin of Brillouin zone $k=0$ will force the lattice Hall viscosity to look more like the continuum Hall viscosity calculated for massive Dirac fermions in 2D.
In addition, starting from the lattice tight-binding model, the continuum limit is derived and shown to yield similar results to previous continuum limit studies~\cite{Taylor2013}.
We have generalized our methods to 3D models and, in particular, we have studied the surface Hall viscosity of 3D TI where the surface Dirac states are gapped as a result of a Zeeman splitting term. Inspired by the expression for the Hall viscosity of single massive 2D Dirac fermion in the continuum, we fit our data and find that the nominally cut-off independent term is much more method independent as well.
The overall feature of our calculations in two or three dimensions is that the Hall viscosity, unlike the Hall conductance, is continuous as one crosses the critical points between topological and trivial phases (or from topological to topological phase).

It is worth noting that the difference between the results of our two methods away from critical points are most likely due to lattice effects and not related to the possible differences between a conventional Hall viscosity and a torsional Hall viscosity as discussed in Refs.~\onlinecite{PhysRevB.91.125303,Hoyos_rev}. The conventional Hall viscosity refers to those calculations in systems which usually have Galilean invariance (including e.g., rotational invariant QHE states and chiral superfluids), and the torsional Hall viscosity refers to those calculations in the Dirac-type models with coupling between the momentum and spin/orbital matrices. The latter requires a coupling to geometry through the frame-fields, while most aspects of the former are satisfactorily treated using just the metric tensor. Both types of calculations share many similar features, but both approaches have yet to be unified into a full understanding.  
In this respect, our calculations (since they reproduce the correct continuum-limit values of both of the potential types of Hall viscosity) suggest that there should be no distinction between the torsional and the conventional Hall viscosities, at least as far as the stress-stress response is concerned. 
This idea is supported by the observation that the continuum limit of our two methods (along with the third method based on the momentum polarization) in the Hofstadter or the Chern insulator models lead to the identical correct results in both cases. Another piece of evidence is based on our calculations for the surface of a 3D TI, where the Hall viscosity computed by the two methods has an almost method independent piece (quadratic in surface gap).

We note that the underlying idea for our derivation of the Hall viscosity is quite general and does not involve any assumption about the type of lattice, crystal symmetry group and the shape of sample. Therefore, our techniques can be adapted to wide variety of lattice configurations and sample geometries.
Remarkably, in the presence of translational symmetry, one may write the Hall viscosity in terms of an integration over lattice momenta which can be evaluated exactly in the thermodynamic limit by means of well-known numerical integration techniques.

Using the electron-phonon coupling (second) approach, we have shown that the Hall viscosity would appear as an anomalous term in the effective action for phonons. This term couples the longitudinal and transverse phononic modes and modifies the dispersion relation. We have exemplified various possible experimental scenarios to observe direct signatures of the Hall viscosity by investigating the phononic properties of a media.

In principle, our methods can be applied to all other interesting gapped or gapless lattice models for interacting or non-interacting topological phases.
Both formulas, Eqs.~(\ref{eq:proj_visc_gauge}) and (\ref{eq:proj_visc_ph}), for the Hall viscosity can be  generalized to the many-body problems
\begin{align*}
\zeta^{(p)}_H=-2\ \text{Im}\  \langle \partial_{u_{12}}\Psi | \partial_{u_{11}}  \Psi\rangle
\end{align*}
where $|\Psi\rangle$ is the many-body wave function. 

Last but not least, provided that the Hall viscosity can be used as another probe along with other topological responses to distinguish topological phases; one can also study the effect of interactions (and maybe disorder) on the viscoelastic behavior using the methods introduced here.

\section{Acknowledgments}
HS would like to thank Maissam Barkeshli, Pouyan Ghaemi and Thomas Tuegel for useful discussions. TLH is supported by NSF (USA), DMR-1351895-CAR. SR has been supported by 
the grant NSF (USA), DMR-1455296 and Alfred P. Sloan foundation.


\begin{appendix}
\renewcommand\theequation{A\arabic{equation}}
\section{Derivation of the Hall viscosity formula for the Chern Insulator}
In order to compute the Hall viscosity using Eq.~(\ref{eq:Kubo_visc}), we need to find the eigenstates of the Hamiltonian, Eq.~(\ref{eq:CI_model}). The Hamiltonian in the reciprocal space reads
\begin{align}
\hat{H}= \sum_{s, \textbf{k}} c_{\textbf{k}}^\dagger h_s(\textbf{k}) \sigma_s c_{\textbf{k}}
\end{align}
where
\begin{align}
h_1(\textbf{k}) &=t_1 \sin k_1 ,\nonumber \\
h_2(\textbf{k})&= t_2 \sin k_2 ,\nonumber \\
h_3(\textbf{k})&= m-r \cos k_1-r \cos k_2.
\end{align}
Here for simplicity, we set (the lattice constant) $a=1$. For each $\textbf{k}$, there are two eigenstates 
\begin{align*}
H |\pm,\textbf{k} \rangle =\pm E_{\textbf{k}} |\pm, \textbf{k}\rangle,
\end{align*}
where
\begin{align*}
|-, \textbf{k}\rangle&= \left( \begin{array}{c}
u_{\textbf{k}} \\ -v_{\textbf{k}} e^{i\theta_{\textbf{k}}}
\end{array} \right), \hspace{1cm}
|+,\textbf{k}\rangle= \left( \begin{array}{c}
v_{\textbf{k}} \\ u_{\textbf{k}} e^{i\theta_{\textbf{k}}}
\end{array} \right)
\end{align*}
and
\begin{align}
E_{\textbf{k}} &= \left( h_1^2(\textbf{k}) + h_2^2(\textbf{k})+h_3^2(\textbf{k})\right)^{1/2},\nonumber \\
u_{\textbf{k}} &= \sqrt{\frac{E_{\textbf{k}}-h_3({\textbf{k}})}{2E_{\textbf{k}}}}, \nonumber \\
v_{\textbf{k}} &= \sqrt{\frac{E_{\textbf{k}}+h_3({\textbf{k}})}{2E_{\textbf{k}}}}, \nonumber \\
\tan \theta_{\textbf{k}}&= \frac{h_2({\textbf{k}})}{h_1({\textbf{k}})} .
\end{align}
The matrix elements of the Pauli matrices in the eigenstate basis are
\begin{subequations}
\begin{align}
\sigma_1^{-+}=\langle -,\textbf{k} | \sigma_1 | +, \textbf{k} \rangle &= \frac{iE_{\textbf{k}} \sin\theta_{\textbf{k}} - h_3 \cos\theta_{\textbf{k}} }{E_{\textbf{k}}} ,\\
\sigma_2^{-+}=\langle -,\textbf{k} | \sigma_2 | +, \textbf{k} \rangle &= \frac{-iE_{\textbf{k}} \cos\theta_{\textbf{k}} - h_3 \sin\theta_{\textbf{k}} }{E_{\textbf{k}}} ,\\
\sigma_3^{-+}=\langle -,\textbf{k} | \sigma_3 | +, \textbf{k} \rangle &= \frac{\sqrt{E_{\textbf{k}}^2  - h_3^2 }}{E_{\textbf{k}}} .
\end{align}
\end{subequations}

\subsection{Minimal coupling method}
We are to calculate
\begin{align*}
\langle -,\textbf{k} | {\cal T}^{(g)}_{11} | +, \textbf{k} \rangle & = (\partial_1h_1 \sigma_1^{-+} + \partial_1 h_3 \sigma_3^{-+}) \sin k_1, \\
\langle -,\textbf{k} | {\cal T}^{(g)}_{12} | +, \textbf{k} \rangle & = (\partial_2 h_2 \sigma_2^{-+} + \partial_2 h_3 \sigma_3^{-+}) \sin k_1.
\end{align*}
So, 
\begin{align}
N_{\textbf{k}}=& \langle -,\textbf{k} | {\cal T}^{(g)}_{11} | +, \textbf{k} \rangle\langle +,\textbf{k} | {\cal T}^{(g)}_{12} | -, \textbf{k} \rangle \nonumber \\ =&{\Big[}  -\frac{\partial_1 h_1 \partial_2 h_2}{E_{\textbf{k}}^2} (h_1h_2 +ih_3 E_{\textbf{k}}) \nonumber \\ &+ \frac{\partial_1 h_3 \partial_2 h_3}{E_{\textbf{k}}^2} (h_1^2+h_2^2) \nonumber \\
&+ \frac{\partial_1 h_1 \partial_2 h_3}{E_{\textbf{k}}^2} (i E_{\textbf{k}} h_2-h_3 h_1)\nonumber \\ 
& + \frac{\partial_1 h_3 \partial_2 h_2}{E_{\textbf{k}}^2} (iE_{\textbf{k}} h_1-h_3 h_2) {\Big]}  \sin^2 k_1
\end{align}
and the Hall viscosity becomes
\begin{align}
\zeta_H^{(g)} &= \frac{1}{L^2} \sum_{\textbf{k}} {\cal B}^{(g)}(\textbf{k})
\end{align}
where the adiabatic curvature is
\begin{align*}
{\cal B}^{(g)}(\textbf{k})=& -2\ \text{Im}{\Big\{} \frac{N_{\textbf{k}}}{(2E_{\textbf{k}})^2}{\Big\}} \nonumber \\
=& \sum_{abc} \epsilon_{abc}\ \frac{h_a \partial_1 h_b \partial_2 h_c}{2E_{\textbf{k}}^3} \sin^2 k_1\nonumber \\
=& \frac{t_1t_2 \sin^2 k_1 (m\cos k_1\cos k_2 - r \cos k_1 - r \cos k_2)}{2E_{\textbf{k}}^3}
\end{align*}
and then we can make it isotropic
\begin{align}
{\cal B}^{(g)}(\textbf{k})=& \frac{t_1t_2 (\sin^2 k_1+\sin^2 k_2)}{4E_{\textbf{k}}^3} \times \nonumber \\
 &(m\cos k_1\cos k_2 - r \cos k_1 - r \cos k_2).
\end{align}

\subsection{Electron-phonon coupling}
For this part, it is more convenient to write everything in terms of $k_1$ and $k_2$ explicitly.
\begin{align*}
\langle -,\textbf{k} | {\cal T}^{(p)}_{11} | +, \textbf{k} \rangle & = t\sin k_1 \sigma_1^{-+} - r\cos k_1 \sigma_3^{-+}, \\
\langle -,\textbf{k} | {\cal T}^{(p)}_{12} | +, \textbf{k} \rangle & = t( \sin k_1 \sigma_2^{-+} + \sin k_2 \sigma_1^{-+}) ,
\end{align*}
hence,
\begin{align}
N_{\textbf{k}}= \langle -,\textbf{k} &| {\cal T}^{(p)}_{11} | +, \textbf{k} \rangle\langle +,\textbf{k} | {\cal T}^{(p)}_{12} | -, \textbf{k} \rangle \nonumber \\ = 
\frac{t}{E_{\textbf{k}}^2} {\Big[}& t \sin^2 k_1 \left(-i h_3 E_{\textbf{k}}- (E_{\textbf{k}}^2 - h_3^2)\sin\theta_{\textbf{k}} \cos\theta_{\textbf{k}}\right) \nonumber \\
&+ t\sin k_1 \sin k_2 (E_{\textbf{k}}^2\sin^2 \theta_{\textbf{k}}+ h_3^2 \cos^2\theta_{\textbf{k}}) \nonumber \\ &-
r \cos k_1 \sin k_1 (i E_{\textbf{k}} h_1 - h_3 h_2 ) \nonumber \\ 
&-r \cos k_1 \sin k_2 (-i E_{\textbf{k}} h_2 - h_3 h_1)
{\Big]} .
\end{align}
Thus the adiabatic curvature is
\begin{align*}
{\cal B}^{(p)}(\textbf{k})= -2\ \text{Im} & {\Big\{} \frac{N_{\textbf{k}}}{(2E_{\textbf{k}})^2}{\Big\}} \nonumber \\ 
=\frac{t}{2E_{\textbf{k}}^3}  {\Big[}  t & \sin^2 k_1 (m-r\cos k_1 -r \cos k_2) \nonumber \\
&+rt \cos k_1 \sin^2 k_1-rt \cos k_1 \sin^2 k_2
{\Big]} \nonumber \\
& \hspace{-1.43cm} = \frac{t^2{\big(} \sin^2 k_1 (m-r\cos k_2)-r \sin^2 k_2 \cos k_1 {\big)} }{2E_{\textbf{k}}^3} ,
\end{align*}
and it can be made isotropic
\begin{align}
{\cal B}^{(p)}(\textbf{k})=&
\frac{t^2 \sin^2 k_1 (m-2r\cos k_2)}{4E_{\textbf{k}}^3} \nonumber \\ 
&+\frac{t^2 \sin^2 k_2 (m-2 r\cos k_1) }{4E_{\textbf{k}}^3}. 
\end{align}

\end{appendix}

\bibliography{HallVisc_v10}

\end{document}